\documentclass[useAMS,usenatbib,usegraphicx]{mn2e}
\usepackage{ gensymb }

\usepackage{graphicx}
\usepackage[fleqn]{amsmath}
\usepackage{amssymb}

\newcommand{\dd}{{\rm d}}
\newcommand{\Pm}{\mathcal{P}}
\newcommand\bnabla{{\bmath\nabla}}
\newcommand\bB{{\bmath B}}

\newcommand\bv{{\bmath v}}
\newcommand\bfI{\mathbf{I}}
\newcommand\bfT{\mathbf{T}}

\newcommand\rmT{\mathrm{T}}
\newcommand\f{\frac}
\newcommand\p{\partial}

\usepackage{epsfig}
\usepackage{amsmath}

\title[Planet migration in weakly magnetised discs]
{Type I planet migration in weakly magnetised laminar discs}

\author[J\'er\^ome Guilet, Cl\'ement Baruteau and John C. B. Papaloizou]
{J\'er\^ome Guilet, Cl\'ement Baruteau and John C. B. Papaloizou\\
Department of Applied Mathematics and Theoretical Physics,
University of Cambridge, Centre for Mathematical Sciences,\\
Wilberforce Road, Cambridge CB3 0WA
}

\begin{document}

\maketitle

\label{firstpage}

\begin{abstract}
The migration of low mass planets, or type I planetary migration, has been studied in hydrodynamical disc models for more than three decades. For a long time, it was thought to be very rapid and directed inwards due to Lindblad torques. More recently, it has been shown that the corotation torque, linked to the horseshoe motion of the gas near the planet, may slow down or even reverse migration. How is this picture modified by the expected presence of a magnetic field in the protoplanetary disc ? When the magnetic field is strong enough to prevent horseshoe motion, the corotation torque is replaced by a torque arising from magnetic resonances which may significantly alter the migration rate. In the case of a weaker magnetic field, the magnetic field is not strong enough to prevent horseshoe motion and a corotation torque then exists. In this regime, recent turbulent MHD simulations have reported the existence of an additional component of the corotation torque due to the presence of the magnetic field. The aim of this paper is to investigate the physical origin and the properties of this additional corotation torque.

We performed MHD simulations of a low mass planet embedded in a 2D laminar disc threaded by a weak toroidal magnetic field, where the effects of turbulence are modelled by a viscosity and a resistivity. We confirm that the interaction between the magnetic field and the horseshoe motion of the gas results in an additional corotation torque on the planet, which we dub the MHD torque excess. We demonstrate that it is caused by the accumulation of the magnetic field along the downstream separatrices of the horseshoe region, which gives rise to an azimuthally asymmetric underdense region at that location. The properties of the MHD torque excess are characterised by varying the slope of the density, temperature and magnetic field profiles, as well as the diffusion coefficients and the strength of the magnetic field. The sign of the torque excess and its radial distribution are found to be in agreement with the earlier full MRI simulations. This sign depends on  the density and temperature gradients only and is positive for profiles expected in protoplanetary discs. The magnitude of the torque excess is in turn mainly determined by the strength of the magnetic field and the turbulent resistivity. It can be strong enough to reverse migration even when the magnetic pressure is less than one percent of the thermal pressure. The MHD torque excess can therefore lead to outward planetary migration in the radiatively efficient outer parts of protoplanetary discs, where the hydrodynamical corotation torque is too weak to prevent fast inward migration. 
\end{abstract}


\section{Introduction}

The gravitational interaction between planets and their parent
protoplanetary disc plays a prominent role in shaping planetary
systems from the early stages of their formation and evolution. The
torque exerted by the disc on a planet drives orbital
migration, a change in the planets semi-major axis during the disc
lifetime. The direction and speed of migration are intimately related
to the planet mass, and the disc properties near the planet's orbital
radius (turbulent viscosity, density and temperature
profiles). Several regimes of migration are commonly distinguished,
based on the planet's ability to open an annular gap around its
orbit. This ability depends on the planet to primary mass ratio, the aspect ratio of the disc $h$, and the viscosity parameter modelling the effect of turbulence, $\alpha$.  For typical disc aspect ratios and alpha
viscosities in regions of planet formation ($h \sim 0.05$, $\alpha
\sim$ a few $\times 10^{-3}$), type I migration applies to planets up
to 10 to 20 Earth masses.  Giant planets in the Jupiter-mass range
carve a deep gap around their orbit and experience type II migration
\citep{lp86}. Sub-giant planets that open a partial gap may undergo
type III runaway migration in massive discs \citep{mp03}.  In this
paper, we focus on type I-migrating planets.

The torque acting on type I-migrating planets has two components: the
differential Lindblad torque and the corotation torque. The
differential Lindblad torque is the rate of angular momentum carried
away by the spiral density waves (wakes) that the planet generates in
the disc. Its sign and magnitude result from a slight asymmetry in the
wakes' density distribution \citep{w97}.  For typical density and
temperature profiles in discs, the differential Lindblad torque is a
negative quantity. Alone, it would make the direction of type I migration inwards, on
a time scale much shorter than the typical lifetime of protoplanetary
discs \citep[e.g.,][]{Tanaka2002}.

The corotation torque, or horseshoe drag, accounts for the angular
momentum exchanged between the planet and the disc material within its
coorbital region. It has been subject to intense investigation after
it was discovered that its amplitude may be comparable to, or even
exceed that of the differential Lindblad torque in radiatively inefficient regions of planet
formation \citep{pm06,
  bm08a, pp08, mc09, Kley09, pbck10, mc10, pbk11}. The corotation
torque is directly related to the distribution of the fluid's
vortensity within the planet's horseshoe region (vortensity refers to
the ratio of the vertical component of the vorticity to the surface
density, and is also known as potential vorticity). In viscous disc
models, the corotation torque is in general the sum of several terms:
\begin{enumerate}
\item A term proportional to the vortensity gradient across the
  horseshoe region, which arises from advection-diffusion of
  vortensity inside this region. It is often called vortensity-related
  corotation torque \citep{wlpi91,masset01}.
\item A term proportional to the entropy gradient across the horseshoe
  region, similarly named entropy-related corotation torque. It
  originates from the (formally) singular production of vortensity at
  the downstream separatrices of the horseshoe region, location at
  which the entropy gets discontinuous due to the advection of entropy
  within the horseshoe region \citep{mc09, pbck10}.  In the limit of
  locally isothermal discs, where the radial profile of temperature is
  constant in time, the corotation torque features instead the local
  temperature gradient \citep{cm09, pbck10}.
\end{enumerate}

The sign and amplitude of the total corotation torque depend on the
disc's density and temperature profiles across the planet's horseshoe
region, as well as on viscosity and thermal diffusivity in this
region. This is in contrast to the differential Lindblad torque, whose
amplitude essentially depends on the local temperature gradient.  The
potentially large amplitude of the corotation torque makes it play a
very important role in the evolution of young planetary systems.  In
this context, much effort has been put forward to produce accurate,
yet simple analytic formulae for the corotation torque \citep{mc10,
  pbk11}.  They may help address some of the shortcomings recently
spotted by models of planet population syntheses \citep{IdaLin4,
  mordasini09b, sli09, HellaryNelson12}.

For type-I migrating planets, the width of the horseshoe region is a
fraction of the disc's pressure scale height, and thus of the typical
size of turbulent eddies. It is {\it a priori} unclear whether
turbulence would generally act as a viscous and thermal diffusion over
the horseshoe region of low-mass planets. Put another way, it is
unclear whether the corotation torque in turbulent discs behaves
similarly as in viscous discs. A few recent studies have started to
examine this issue \citep{bl10, bfnm11, Uribe11,pierens12}. Adopting the turbulence model
originally developed by \cite{lsa04}, where turbulence is generated by
stochastic forcing, the 2D hydrodynamical simulations of \cite{bl10} and \cite{pierens12}
showed that both the differential Lindblad torque and the corotation
torque behave similarly as in equivalent laminar viscous
discs. 

How is the picture modified by the expected presence of a magnetic field ? In the case of a strong azimuthal magnetic field that prevents horseshoe motion, the corotation torque is replaced by angular momentum carried away at magnetic resonances, where the rotational velocity relative to the planet is equal to the propagation speed of a slow MHD wave or an Alfv\'en wave \citep{terquem03,fromang05}. In the intermediate case of a weak magnetic field, horseshoe motion can be expected to take place and give rise to a corotation torque. This has been confirmed by simulations of discs in which the magneto-rotational instability (MRI) drives magneto-hydrodynamical (MHD) turbulence. \cite{bfnm11} (hereafter BFNM11) carried out 3D MHD simulations of discs with turbulence due to the MRI operating throughout the whole disc. They adopted a disc model with a weak mean toroidal magnetic field, in
which non-ideal MHD effects and vertical stratification were
neglected.  They showed the existence of horseshoe dynamics and an
unsaturated corotation torque for Saturn-like planets embedded in
thick ($h = 0.1$) discs. They also found the existence of an
additional corotation torque with moderate amplitude.  \cite{Uribe11}
performed 3D MHD simulations with a similar setup as in \cite{bfnm11},
except for the inclusion of vertical stratification. Their simulations
with intermediate-mass planets showed outward type I migration, which
is indicative of the presence of a fairly large and unsaturated
positive corotation torque in their simulations.

The aim of this paper is to investigate the physical origin and the properties of this additional corotation torque due to a weak magnetic field, discovered by MHD turbulent simulations. For this purpose, we adopt a two-dimensional laminar disc model threaded by a weak toroidal magnetic field, and in which turbulence is modelled by viscous and magnetic diffusion. This simplified setup is chosen for several purposes. First, the lower computational cost of performing 2D simulations allows a wide exploration of the parameter space, so far inaccessible to 3D turbulent simulations. Second, the simplicity of the model allows a deeper physical understanding, which should prove useful to interpret more complex turbulent simulations. This approach is not new but has been carried out in several other contexts (see Section~\ref{sec:setup} below). Modelling the effects of turbulence by diffusion coefficients is obviously a strong assumption, which should be checked for the problem considered in this paper by comparing with numerical simulations of turbulent discs (a first step in this direction is done in Section~5 of this paper). We note, however, that such an approach has been successful in studies of migration in hydrodynamical disc models as mentioned earlier in this introduction.

Our simulation results confirm the existence of an additional corotation torque acting on low-mass
planets embedded in discs with a toroidal magnetic field. A wide
exploration of the parameter space allows us to determine numerically
how the sign and magnitude of this additional corotation torque depend
on the disc properties (radial profiles of density, temperature and
toroidal magnetic field, viscosity, magnetic diffusivity). For typical
density and temperature profiles, the additional corotation torque is
positive. Its amplitude has a very steep dependence on the magnetic field strength and diffusivity, and in some cases it may largely exceed the
amplitude of the differential Lindblad torque.

This paper is organised as follows. The physical model and numerical
setup are presented in Section~\ref{sec:setup}. The topology of the magnetic
field within the horseshoe region is described in Section~\ref{sec:magnetic}. The
properties (sign, magnitude) of the new corotation torque in weakly
magnetised discs are examined in Section~\ref{sec:torque}. A detailed
comparison with the results of the 3D MHD turbulent simulations of BFNM11 follows in Section~\ref{sec:turblike}.  Concluding remarks and future directions are given in Section~\ref{sec:conclusion}.

\section{Physical model and numerical setup}
	\label{sec:setup}
\subsection{Disc model}
	\label{sec:setup_disc}
We consider a two-dimensional disc model, assuming a locally isothermal equation of state (i.e. the temperature depends on radius but not on time). The gas is assumed to be non-selfgravitating. The disc is laminar, and the effects of MHD turbulence are modelled by effective diffusion coefficients: a viscosity and a resistivity. Such an approach has been used in the past in a variety of contexts. Since the seminal work of \citet{shakura73}, hydrodynamical disc models with an effective turbulent viscosity have been widely used to study many different aspects of disc dynamics, including the problem of planet migration. An effective resistivity has been used for the study of magnetic flux transport in accretion discs \citep[e.g.][]{lubow94a,guilet12} and the launching of jets \citep[e.g.][]{ferreira95,zanni07}. The viscosity is parameterised by the usual $\alpha$ parameter : $\nu = \alpha c_s^2/\Omega_K$, where $c_s$ is the sound speed and $\Omega_K$ is the keplerian angular velocity. The value of the magnetic diffusivity $\eta$ is then determined by the assumed magnetic Prandtl number defined as $\Pm = \nu/\eta$, giving : $\eta = \alpha/\Pm c_s^2/\Omega_K$. The effective resistivity of MHD turbulence driven by the MRI has been measured in numerical simulations of local disc models \citep{guan09,lesur09,fromang09}. They showed that the magnetic Prandtl number of MHD turbulence is of order unity, though it could differ from 1 by a factor of a few depending on the magnetic field configuration. In most of this article, we therefore assume $\Pm=1$, and relax this assumption in Sections~\ref{sec:alpha_beta} and \ref{sec:turblike}. As for the radial dependence of the diffusion coefficients, either uniform alpha parameter or uniform diffusion coefficients (i.e. $\nu$ and $\eta$) are considered, giving essentially the same results if the value at the planet's orbital radius is the same.

The basic equations solved are :
\begin{equation}
  \f{\p\rho}{\p t}+\bnabla\cdot(\rho\bv)=0,
\end{equation}
\begin{eqnarray}
  \lefteqn{\rho\left(\f{\p\bv}{\p t}+\bv\cdot\bnabla\bv\right)=-\rho\bnabla\Phi-\bnabla p}&\nonumber\\
  &&+\f{1}{\mu_0}(\bnabla\times\bB)\times\bB+\bnabla\cdot\bfT,
\end{eqnarray}
\begin{equation}
  \f{\p\bB}{\p t}=\bnabla\times(\bv\times\bB-\eta\bnabla\times\bB),
\end{equation}
\begin{equation}
  \bfT=\rho\nu\left[\bnabla\bv+(\bnabla\bv)^\rmT-\f{2}{3}(\bnabla\cdot\bv)\,\bfI\right],
\end{equation}
where $\rho$ is the density, $\bv$ the velocity, $\Phi$ the
gravitational potential, $p$ the pressure, $\bB$ the magnetic field,
$\bfT$ the viscous stress tensor, $\eta$ the magnetic diffusivity, $\nu$ the
kinematic viscosity and $\bfI$ the unit tensor of second rank.

A cylindrical polar coordinate system $(r,\varphi,z)$ is used, the disc occupying the plane $z=0$. We assume a simple initial magnetic field geometry, with only a toroidal component $B_\varphi$. This is motivated by the fact that in sheared MHD turbulence the azimuthal component of the magnetic field tends to be the strongest one. Furthermore, we initialise the calculation with power law profiles for the surface density $\Sigma$, temperature $T$ and azimuthal magnetic field $B_\varphi$. The power law indices are defined as : 
\begin{equation}
b \equiv \frac{d\log B_\varphi}{d\log r},
	\label{eq:def_b}
\end{equation}
\begin{equation}
p \equiv \frac{d\log\Sigma}{d\log r},
	\label{eq:def_p}
\end{equation}
\begin{equation}
q \equiv \frac{d\log T}{d\log r}.
	\label{eq:def_q}
 \end{equation}
 
 The strength of the magnetic field is measured in terms of the plasma parameter $\beta$, which is defined as the ratio of the thermal to the magnetic pressure:
  \begin{equation}
  \beta = P_{\rm th} / P_{\rm mag} = 2c_s^2 / v_A^2,
  	\label{eq:def_beta}
  \end{equation}
where $v_A\equiv B/\sqrt{\mu_0\rho}$ is the Alfv\'en velocity, and $\mu_0$ is the vacuum permeability.
 
The azimuthal velocity is initialised with a profile corresponding to a balance between the gravity of the star, the centrifugal force, the pressure gradient and the magnetic force. In can be written (in a non rotating frame):
 \begin{equation}
 v_\varphi = r\Omega_K\left[ 1 + \left(p + q + \frac{2(b+1)}{\beta} \right)h^2 \right],
 \end{equation}
 where $\Omega_K\equiv\sqrt{GM_*/r^3}$ is the Keplerian angular frequency, $h\equiv H/r$ is the aspect ratio of the disc with $H=c_s/\Omega_K$ the disc scaleheight.

 \subsection{Planet}
 A planet is introduced in the disc at the beginning of the simulations and is held in a fixed circular orbit, at a radius $r_p$ and azimuth $\varphi_p=0$.  We work in the frame rotating with the planet at angular frequency $\Omega_p$ and therefore include the required indirect terms in the equation of motion. The planet potential is smoothed over a softening length $\varepsilon = 0.6H(r_p)$ (though this fiducial value is varied in Section~\ref{sec:softening}) and is expressed as: 
\begin{equation}
\Phi_p=-\frac{GM_p}{\left(|\vec{r}-\vec{r_p}|^2 + \epsilon^2\right)^{1/2}}.
\end{equation}
 
 The torque exerted by the disc on the planet is computed with the classical expression :
 \begin{equation}
 \Gamma = \int\int \Sigma \frac{\partial \Phi_p}{\partial \varphi}r\dd r \dd \varphi
 \end{equation}
Unless otherwise specified, the torque calculations presented in this paper include the planet's Hill sphere, whose radius is:
\begin{equation}
 r_H \equiv r_p\left( \frac{M_p}{3M_*}\right)^{1/3}.
 	\label{eq:hill}
 \end{equation}
The effect of excluding the Hill sphere is studied in Section~\ref{sec:softening}. The Hill sphere is also excluded in Section~\ref{sec:turblike} for the comparison with BFNM11. Finally, note that all the torques and torque density distributions presented in this paper are per unit mass of the planet, although we then refer to it as "torque" rather than "specific torque" for conciseness.

 \subsection{Normalisation, fiducial parameters and migration regime}
 	\label{sec:fiducial}
Without loss of generality, we normalise the radius, surface density and time such that $r_p = 1$, $\Sigma(r_p) = 1$, and $\Omega_p=1$. The magnetic field is then normalised such that $\mu_0=1$. The free parameters of the problem are then the planet-to-primary mass ratio $M_p/M_*$, the value at the planet orbital radius of the aspect ratio of the disc $h$, the plasma parameter $\beta$, and the diffusion coefficients (determined by $\alpha$ and $\Pm$), and the slopes indices of the magnetic field $b$, the surface density $p$, and temperature $q$. We use as fiducial parameters the following values: $M_p = 2\times 10^{-5} M_{\star}$ (around 7 earth mass's for a Sun-like star), $h = 0.05$, $\beta=100$, $\alpha=5.10^{-3}$, $\Pm=1$, $b=-1$. For $p$ and $q$, we consider two fiducial disc models (the same ones as in BFNM11) : model~1 has $p=-1/2$ and $q=-1$ (corresponding to a uniform aspect ratio of the disc). Model~2 has $p=-3/2$ and $q=0$, which corresponds to the special case where the corotation torque vanishes in a non magnetised disc, because the vortensity and the temperature are uniform. Most of the parameters are then varied around this fiducial set of values. 

\paragraph*{Migration regime.} The fiducial value of the planet mass falls into the regime of type I migration, since $M_p/(M_*h^3)=0.16$. The relative strength of the magnetic field and horseshoe motion may be measured with the ratio of the Alfv\'en speed to the shear velocity at the separatrix of the horseshoe region. With the half-width of the horseshoe region in a hydrodynamical case estimated as: $x_s \simeq 1.1r_p\sqrt{M_p/(M_*h)}$, one obtains:
\begin{equation}
\frac{v_A}{v_\varphi(r_p-x_s)} \simeq 0.86 \sqrt{\frac{M_*h^3}{M_p\beta}}. 
	\label{eq:va_vshear}
\end{equation}
The fiducial parameters give: $v_A/v_\phi(x_s) \simeq 0.21$. Since the Alfv\'en speed is significantly smaller than the shear velocity at the separatrix of the horseshoe region, the magnetic field is not expected to prevent horseshoe motion of the gas. The numerical results indeed show the presence of horseshoe dynamics.

\paragraph*{Magnetic resonances.} 
In the presence of an azimuthal magnetic field, \citet{terquem03} and \citet{fromang05} showed that magnetic resonances can arise at the locations where the shear velocity equals the propagation velocity of slow MHD waves (one resonance on each side of the planet). If the shear is the unperturbed Keplerian shear, then the separation between the two magnetic resonances (i.e. the width of the region where the flow is sub-slow MHD) can be expressed as:
\begin{equation}
\Delta r_{mr} = \frac{4H}{3\sqrt{1+\beta/2}}.
	\label{eq:magnetic_resonances}
\end{equation}
With our fiducial parameters, this gives $\Delta r_{rm}=9.3\times 10^{-3}$. In the highest resolution runs of the convergence study presented in the Appendix, this region is resolved by $6-7$ cells. Yet, we do not see any evidence for magnetic resonances. The very good convergence of the results shows that numerical diffusion plays a minor role compared to physical diffusion, and suggests that the magnetic resonances are absent for a physical reason rather than a numerical reason. 

Two physical processes may be the cause for the absence of magnetic resonances in our setup: applied resistive diffusion and horseshoe dynamics. Since the postulated magnetic resonances are close to the corotation radius for our fiducial magnetic field strength, diffusion may play an important role. The time to diffuse from one magnetic resonance to the other is $\Delta r_{mr}^2/\eta $, during which a fluid element on a magnetic resonance moves by $\Delta \phi = 16/9\Pm/\alpha(1+\beta/2)^{-3/2}h \simeq 0.05 \simeq h$ for our fiducial parameters. Given this small displacement during a diffusion time, diffusion is expected to have a significant impact on the dynamics at the magnetic resonances.

If they were present, the postulated magnetic resonances would lie within the horseshoe region. The shear is, however, radically modified inside the horseshoe region, thus changing the conditions for the magnetic resonances to appear. The shear velocity can be expected to equal the slow MHD speed in a small region close to the stagnation point, however it is unclear whether slow MHD waves can be launched in these conditions, which are very different from those in the absence of horseshoe motion. In conclusion, because of these two physical reasons we may expect horseshoe motion slightly modified by the presence of the magnetic field rather than magnetic resonances, which is confirmed by the numerical results.

\paragraph*{Timescales.} The properties of the corotation torque are sensitive to the ratio of the diffusion timescale to the u-turn or libration timescales. The diffusion timescale across the horseshoe region is: 
\begin{equation}
\tau_{\rm diff} \equiv \frac{x_s^2}{\nu} \simeq 1.21\frac{M_p}{M_*h^3\alpha}\Omega_p^{-1},
	\label{eq:tdiff}
\end{equation}
which gives $\tau_{\rm diff} \simeq 6.2$ orbits at the planet's orbital radius for the fiducial parameters. The libration timescale is:
\begin{equation}
\tau_{\rm lib} \equiv \frac{8\pi r_p}{3\Omega_p x_s} \simeq \frac{8\pi}{3.3}\sqrt{\frac{M_*h}{M_p}}\Omega_p^{-1},
	\label{eq:tlib}
\end{equation}
giving $\tau_{\rm lib} \simeq 61$ orbits for the fiducial parameters. And finally, the U-turn timescale is approximately $\tau_{\rm U-turn} \simeq h\tau_{\rm lib}$, giving $\tau_{\rm U-turn} \simeq 3$ orbits for our fiducial parameters. We therefore have the ordering $\tau_{\rm U-turn}< \tau_{\rm diff}<\tau_{\rm lib}/2$, which ensures that the hydrodynamical corotation torque remains close to its maximum fully unsaturated value.

 \subsection{Numerical method}
\begin{figure*} 
  \centering
    \includegraphics[width=\columnwidth]{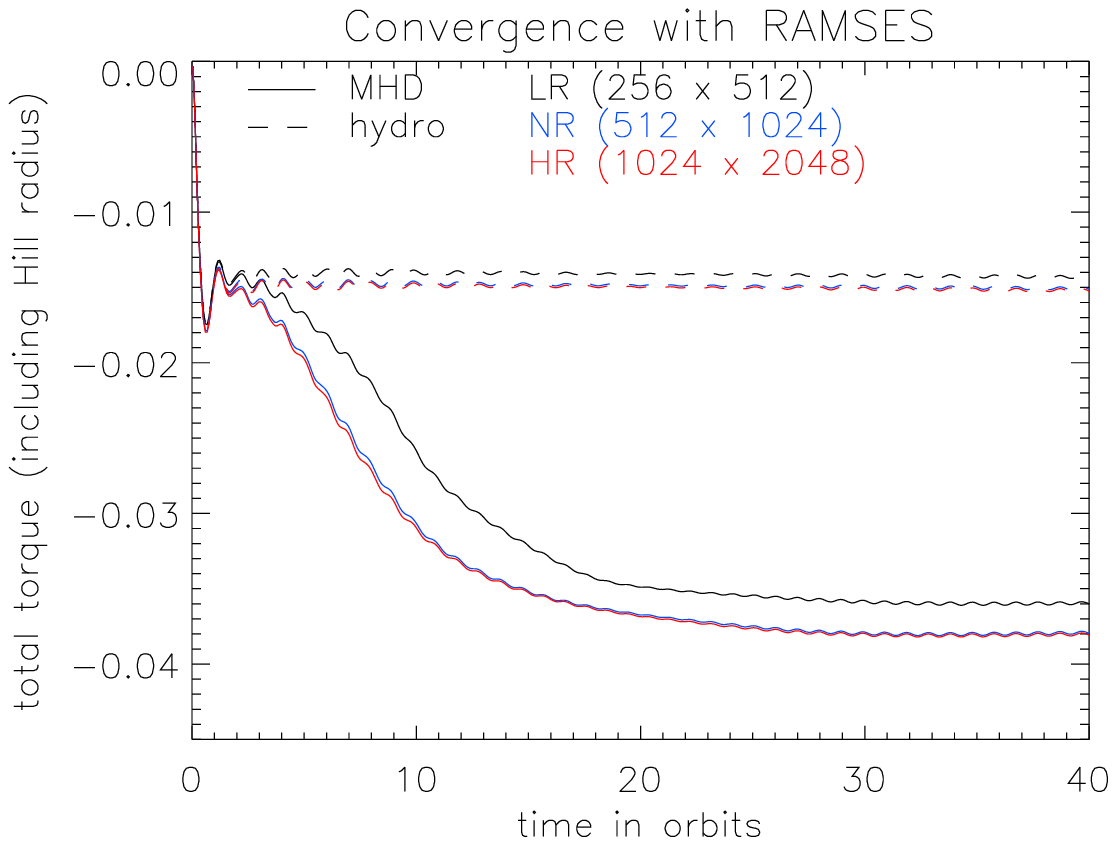}
    \includegraphics[width=\columnwidth]{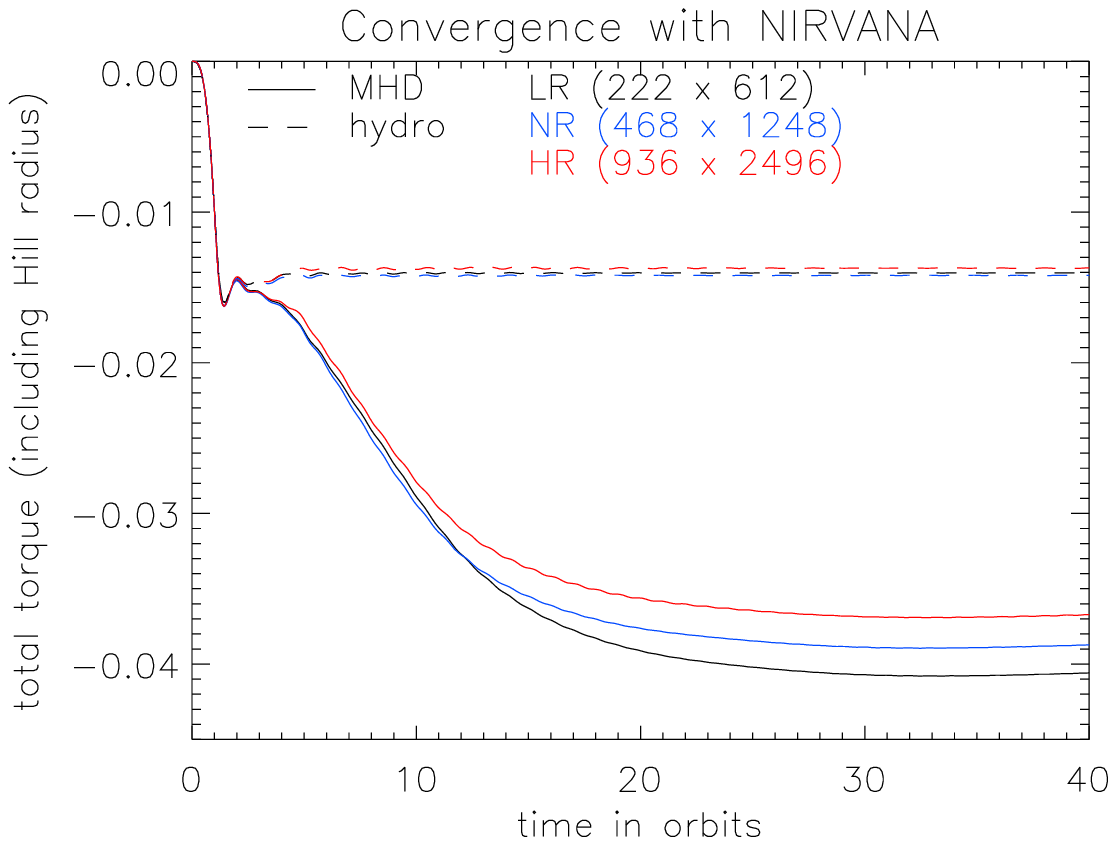} 
  \includegraphics[width=\columnwidth]{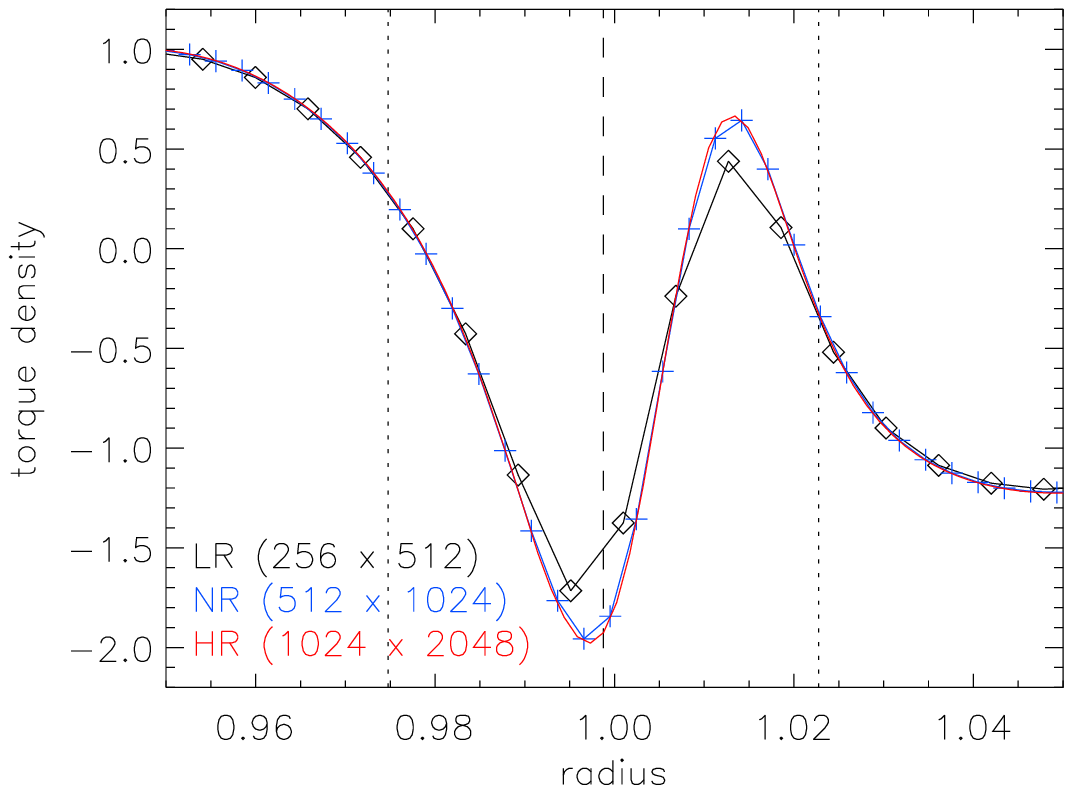}
  \includegraphics[width=\columnwidth]{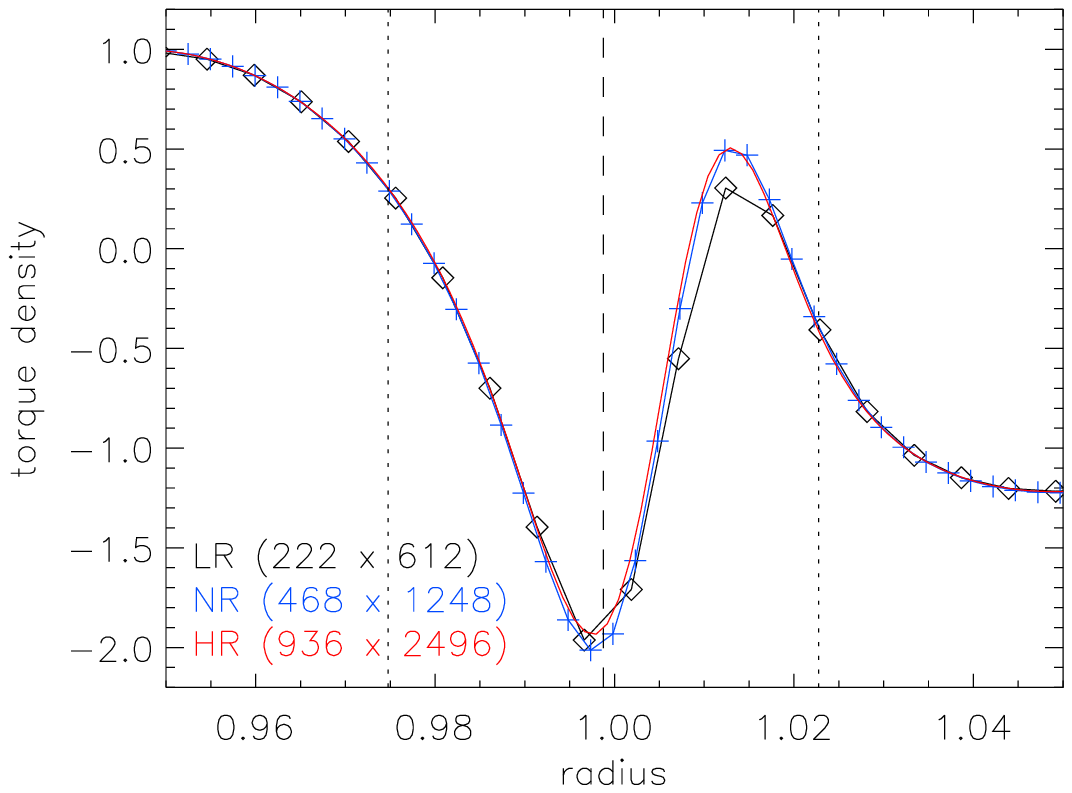}
  \caption{Convergence test of the two codes RAMSES (left panels) and NIRVANA (right panels) for the model 2 ($p=-3/2$ and $q=0$) with our fiducial parameters. The upper panels show the total torque exerted on the planet, MHD simulations being represented with full lines and hydrodynamical ones with dashed lines. The lower panels show the torque density distribution as a function of radius near the planet for the MHD simulations (normalised by the maximum value of the high resolution run). The vertical dashed line shows the corotation radius, while the dotted lines show the extent of the horseshoe region. Three resolutions have been used: the nominal resolution used in this paper (NR, blue curves: $512\times1024$ for RAMSES and $468\times1248$ for NIRVANA runs), a twice lower resolution (LR, black curves: $256\times512$ for RAMSES and $222\times612$ for NIRVANA runs), and a twice higher resolution (HR, red curves: $1024\times2048$ for RAMSES and $936\times2496$ for NIRVANA runs). Nominal and high resolution are in excellent agreement for the RAMSES runs, and reasonable agreement for the NIRVANA runs. The torques computed with both codes at nominal or high resolutions agree within $\simeq 5\%$. }
  	\label{fig:convergence}
\end{figure*}

We use two different MHD codes : the finite volume code RAMSES \citep{teyssier02,fromang06} using the MUSCL-Hancok Godunov Scheme, and the finite difference code NIRVANA \citep{ziegler97} which uses an algorithm similar to the ZEUS code \citep{stone92}. The results of the two codes are shown to be in good agreement in Appendix~A. Figures~\ref{fig:cart} to \ref{fig:drho}, \ref{fig:torque_time_q_diff}, \ref{fig:p-1/2_vary_q}, and \ref{fig:stability} to \ref{fig:phistagn_beta} were done using the code RAMSES. Figures~\ref{fig:B_profiles} and \ref{fig:p-3/2_vary_q} were done using NIRVANA. Finally, Figures~\ref{fig:p-3/2_vary_q}, \ref{fig:turblike} and \ref{fig:convergence} were done using both codes.

The numerical domain extends around the planet location in the range $r\in[0.5,2]$, and $\phi\in[-\pi,\pi]$. Wave-killing zones are used near the grid's inner edge ($r\in[0.5,0.65]$) and outer edge ($r\in[1.7,2]$) in order to avoid spurious reflections of the planet's wakes. For this purpose the density and velocity are damped to their initial values. The magnetic field however is not damped, because the constraint that its divergence remains zero would require a special treatment. We found that this was enough to avoid any significant refection at the edges  where reflective boundary conditions are applied (see Figure~\ref{fig:cart}). Our default grid resolution is $n_r=512$, $n_\phi=1024$ with RAMSES, and $n_r=468$, $n_\phi=1248$ with NIRVANA. It is such that the half-width of the horseshoe region is resolved by about 8 cells for our fiducial set of parameters. In Appendix~A, we show that this resolution is good enough to obtain converged results.

\begin{figure*} 
  \centering
    \includegraphics[width=\columnwidth]{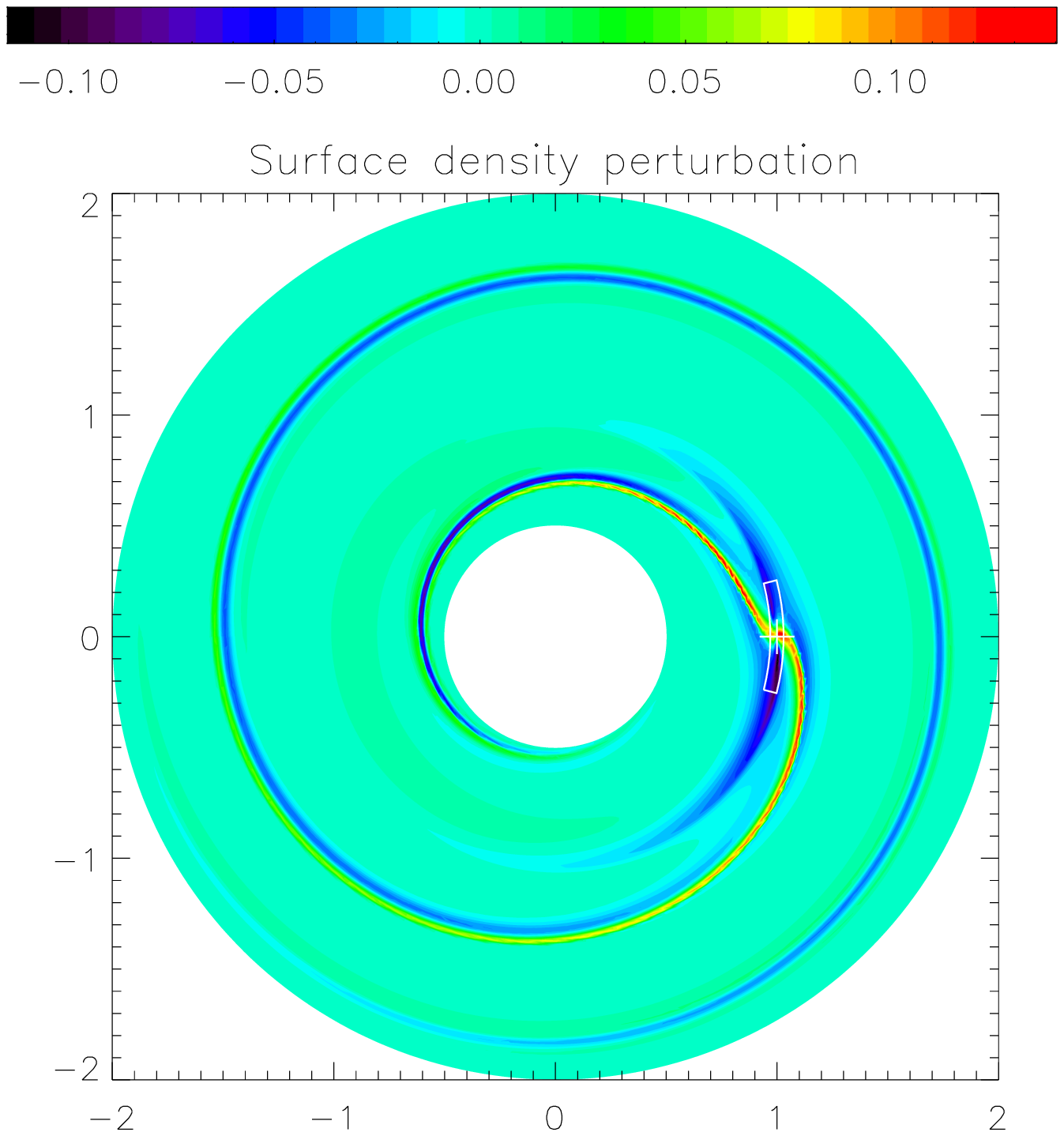}
    \includegraphics[width=\columnwidth]{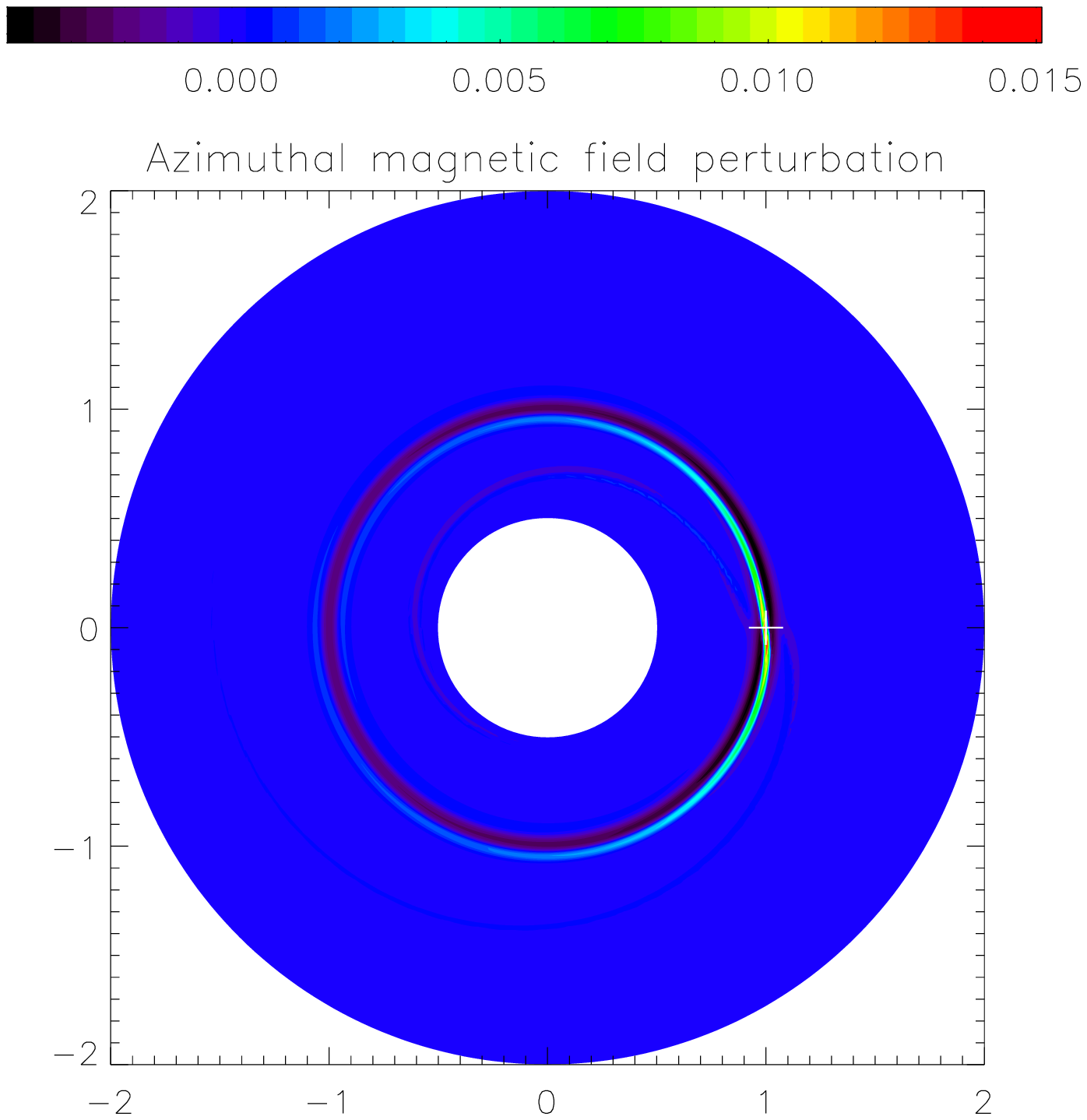}
	  \caption{Global view of a simulation of Model~1 ($p=-1/2$, $q=-1$) with the fiducial parameters. The left panel shows the surface density perturbation, while the right panel shows the perturbation of the azimuthal component of the magnetic field. The location of the planet is shown with a white + sign at $r=1$, $\varphi=0$. The extent of the region around the planet represented in Figures~\ref{fig:B_field_model1}, \ref{fig:B_field_model2} and \ref{fig:drho} is shown with a white sector in the left panel.}
  \label{fig:cart}
\end{figure*}

\section{Magnetic field distribution inside the planet's horseshoe region}
	\label{sec:magnetic}

\begin{figure*}
   \centering    
   \includegraphics[width=0.66\columnwidth]{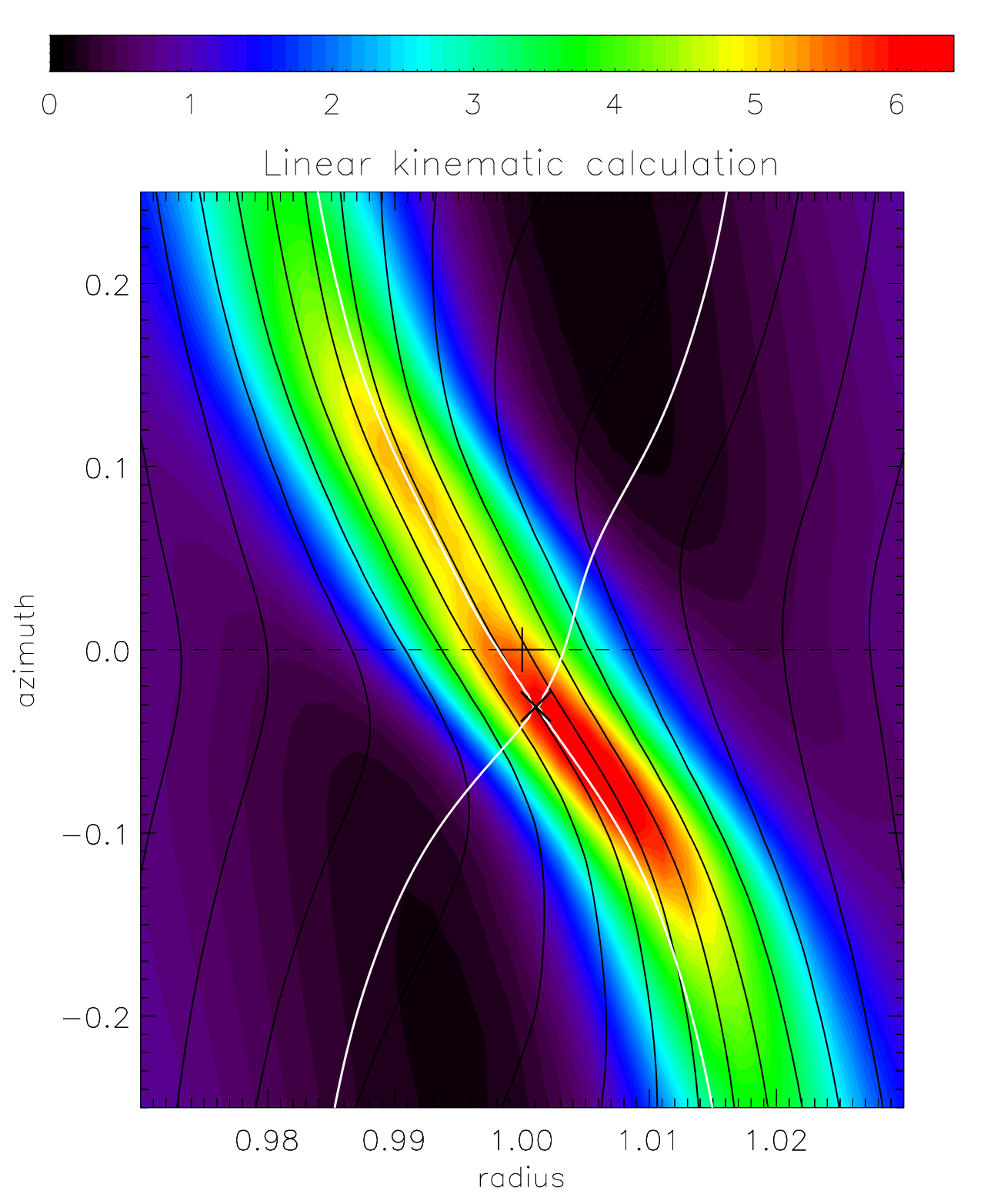}
    \includegraphics[width=0.66\columnwidth]{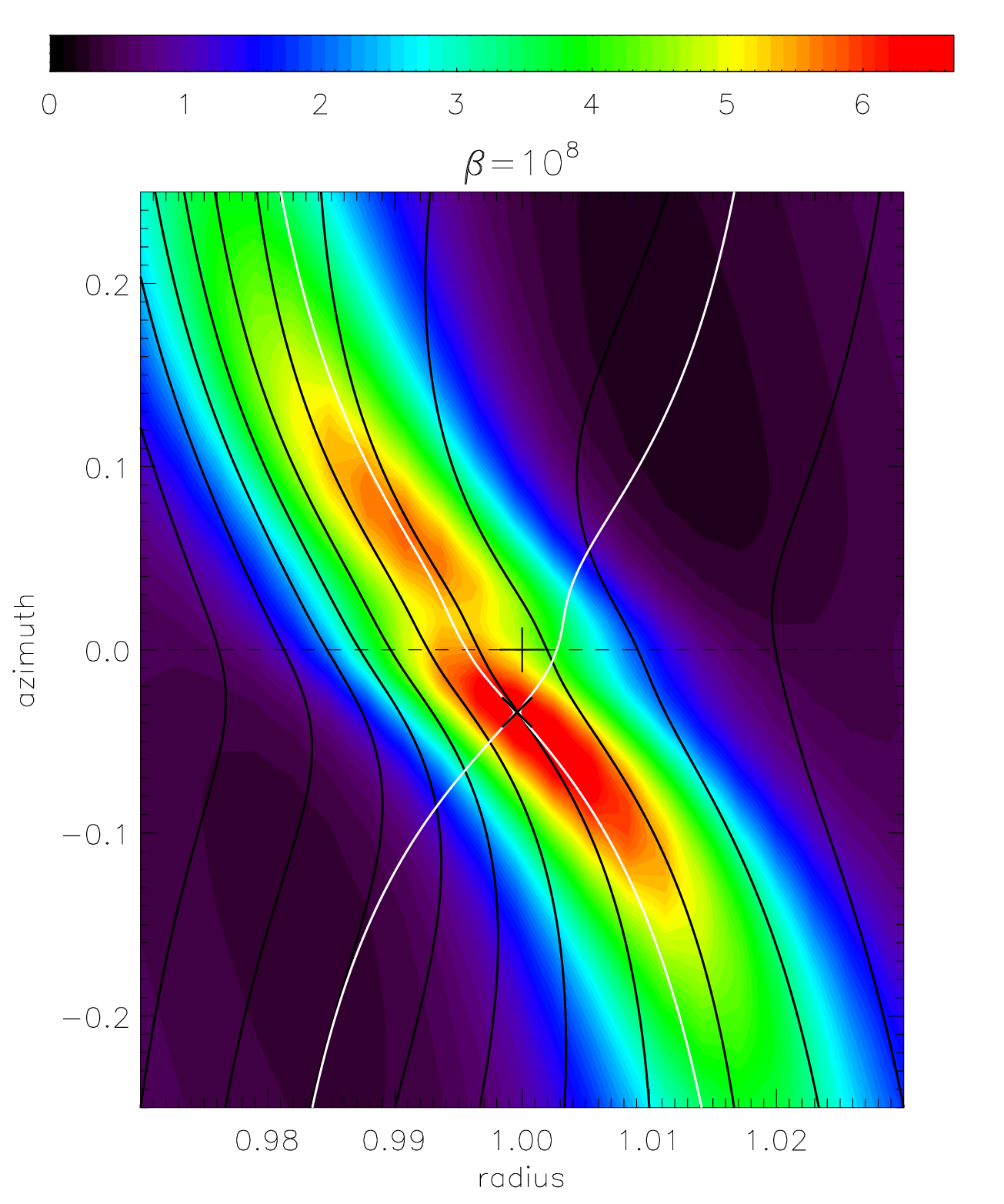}
    \includegraphics[width=0.66\columnwidth]{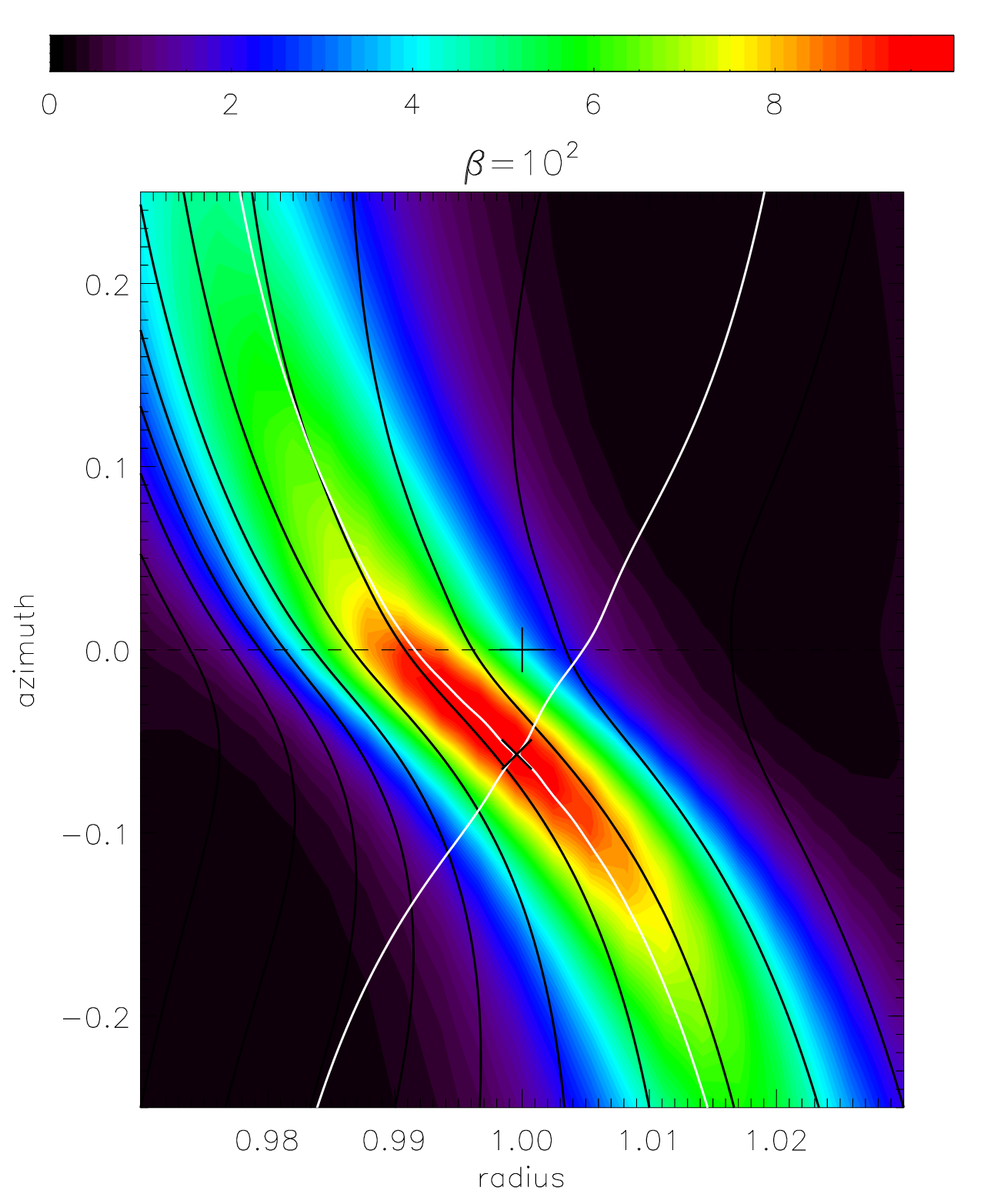}
   \caption{Magnetic field configuration in model 1 ($p=-1/2$, $q=-1$). Left panel: linear hydrodynamical calculation coupled with an advection-diffusion equation for a passive magnetic field. Middle panel: MHD simulation with a very weak magnetic field ($\beta=10^8$). Right panel: MHD simulation with our fiducial magnetic field strength ($\beta=100$). The colour contours show the magnetic energy normalised by its initial value. The black lines represent magnetic field lines, while the white lines represent the separatrices delimiting the horseshoe region. The black cross shows the position of the stagnation point at the intersection of the two separatrices. The planet location is depicted by a black + sign at $r=1$,$\varphi=0$.}
   \label{fig:B_field_model1}
\end{figure*}
 
\begin{figure*}
   \centering
    \includegraphics[width=0.66\columnwidth]{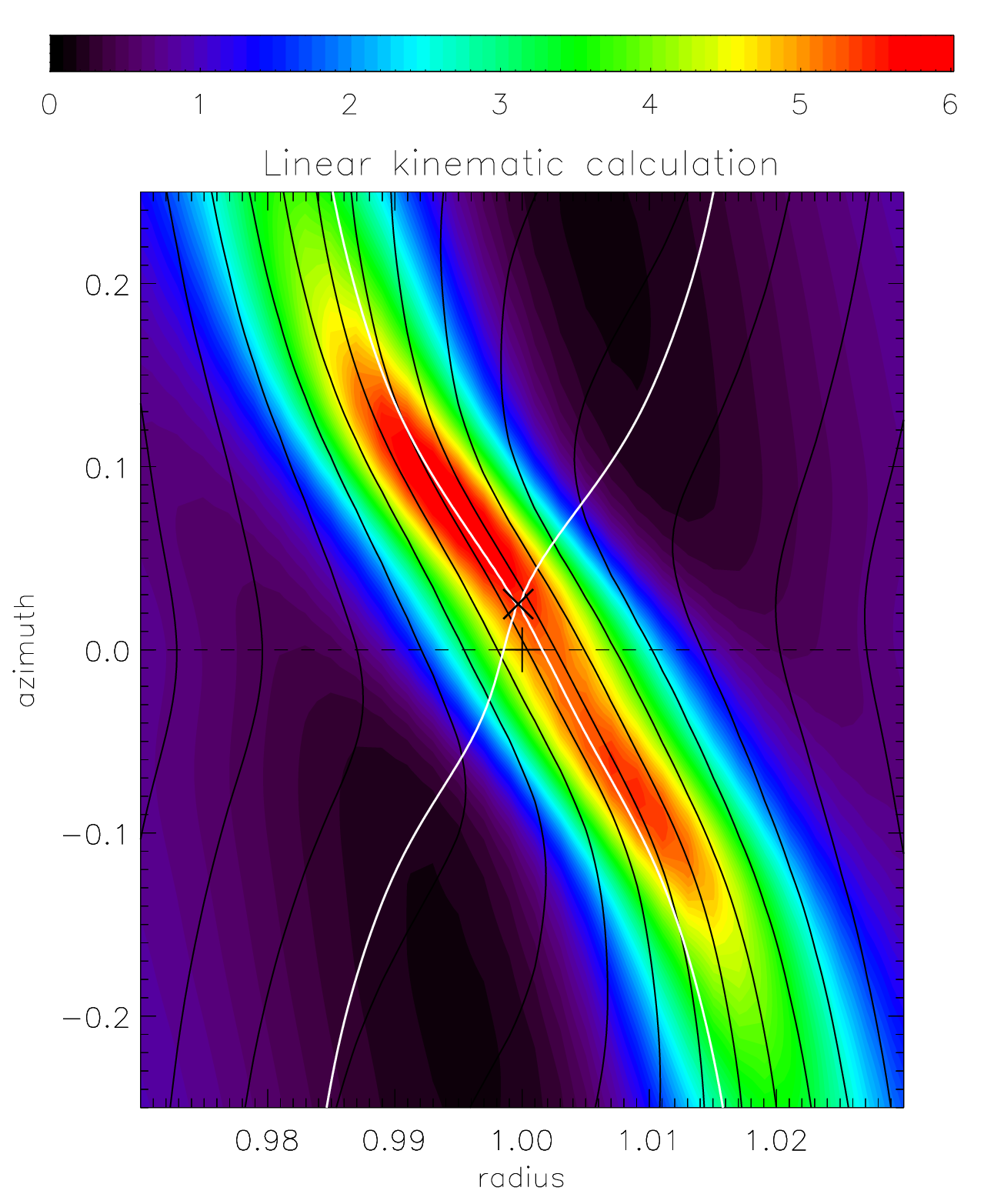}
    \includegraphics[width=0.66\columnwidth]{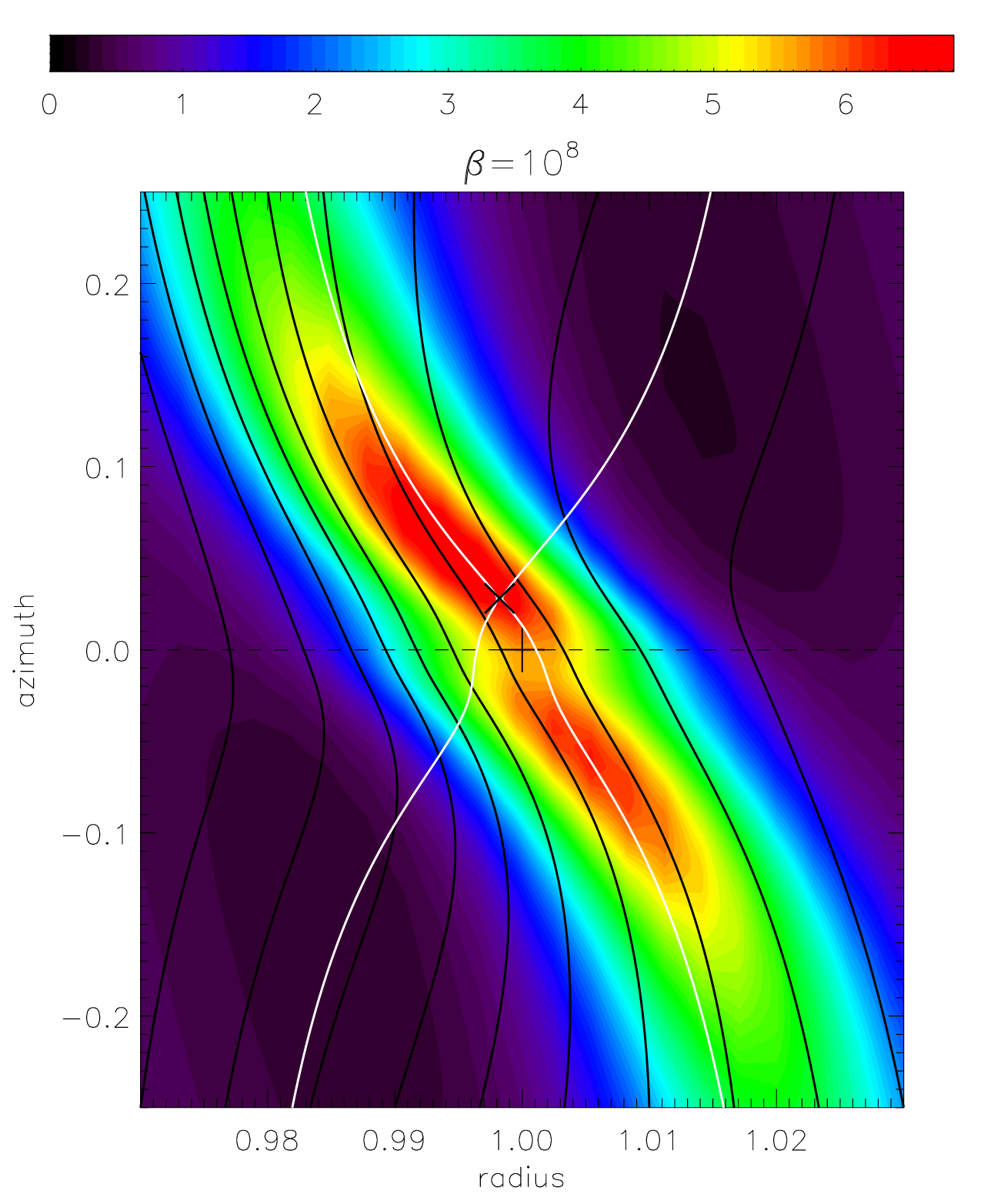}
    \includegraphics[width=0.66\columnwidth]{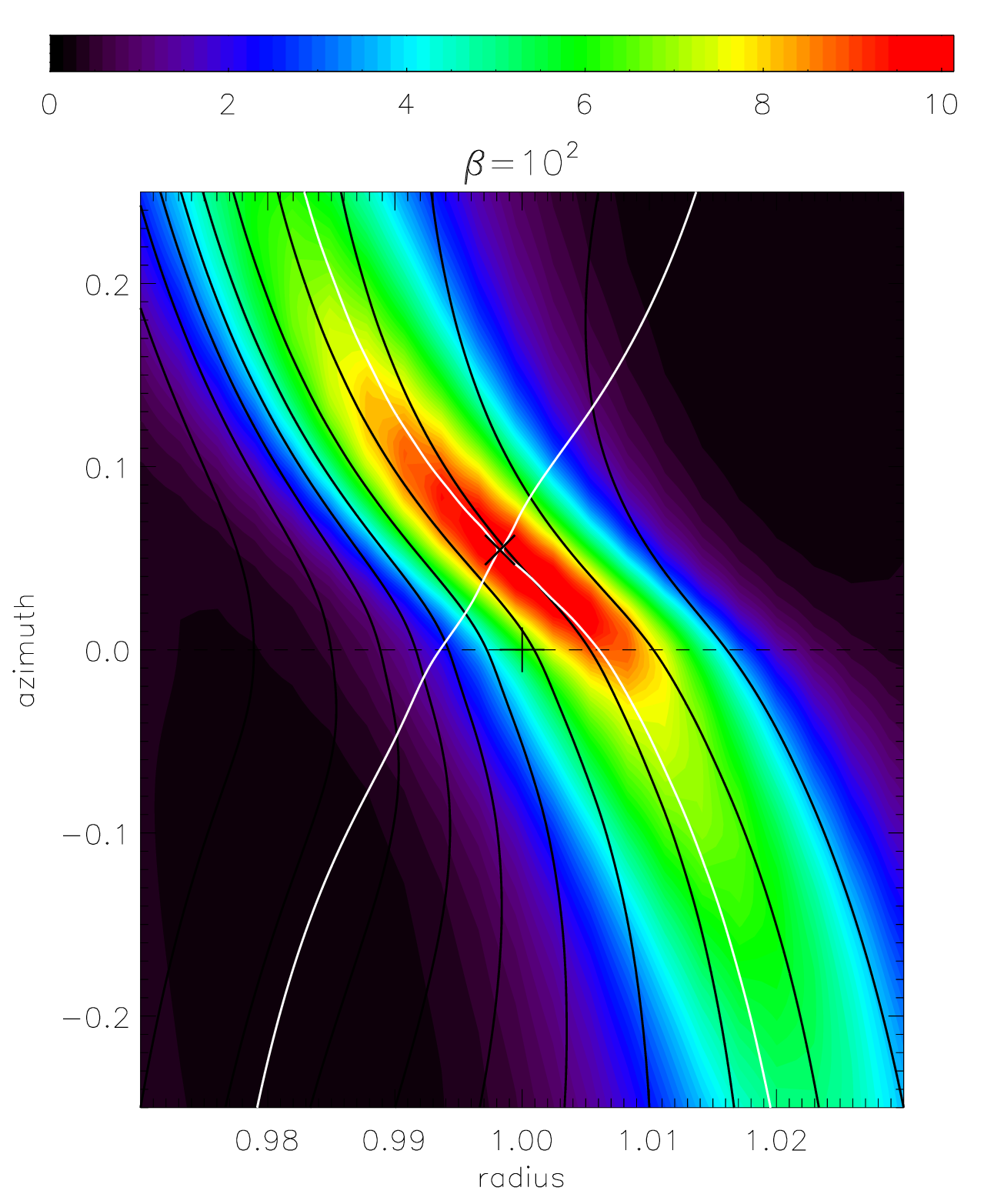}
     \caption{Same as Figure~\ref{fig:B_field_model1} but for model~2 ($p=-3/2$, $q=0$).}
   \label{fig:B_field_model2}
\end{figure*}	

A global view of a simulation of Model~1 with the fiducial parameters is represented in Figure~\ref{fig:cart}. The density perturbation (left panel) shows two distinct features: the usual wake (i.e. the spiral density waves launched in the inner and the outer disc), and an underdense region localised around the planet's orbital radius. This second feature is due to the horseshoe dynamics in the presence of a magnetic field and will be described further in Section~\ref{sec:torque_model1-2}. The magnetic field perturbation shows also two features related to the wake and the horseshoe dynamics, but the latter is clearly the most prominent one. This highlights the fact that the main effect of the relatively weak magnetic field considered here is on the horseshoe dynamics.

We continue the description of the numerical results by considering the effect of the horseshoe gas motion on the magnetic field configuration, as is illustrated by Figures~\ref{fig:B_field_model1} for Model~1 ($\Sigma \propto r^{-1/2}$, $T \propto r^{-1}$), and by Figure~\ref{fig:B_field_model2} for Model~2 ($\Sigma \propto r^{-3/2}$, uniform temperature). We compare simulations using our fiducial magnetisation $\beta=100$ (right panels) with simulations using an extremely weak magnetisation ($\beta=10^8$, middle panels) where the magnetic field is effectively passive. This allows to disentangle the magnetic field perturbation due to a pure advection/diffusion of the magnetic field from its dynamical effects on the velocity field. The extent of the horseshoe region is shown by the separatrices overplotted with white lines: as expected the magnetic field is not strong enough to prevent horseshoe motion of the gas. The magnetic configuration (illustrated by the magnetic pressure with the colour contours and the magnetic field lines in black) is quite similar in the passive and non-passive field cases: the magnetic field lines are bent by the horseshoe motion of the gas, resulting in an accumulation of magnetic flux along the downstream separatrices and in particular close to the stagnation point. The width of the region over which the magnetic flux accumulates is set by an equilibrium between the advection due to horseshoe motion bringing magnetic flux toward the downstream separatrix and the diffusion of this magnetic flux by the action of resistivity. Decreasing the resistivity results in a narrower region of accumulation, and therefore a stronger amplification of the magnetic field there (not shown in the figures because the pattern is very similar). We note that the presence of a non-negligible magnetic field induces slightly more compression of the magnetic field near the planet. This may be linked to a (rather surprising) slight increase of the width of the horseshoe region due to the presence of the magnetic field  (see Figure~\ref{fig:phistagn_beta}). Indeed a wider horseshoe region results in a larger shear velocity at the horseshoe separatrix and a larger advection velocity in the horseshoe region, making the advection process (and therefore the magnetic field compression) more efficient.

Another point which will be of great importance in our discussion of the torque in Section~\ref{sec:torque} is the asymmetry of the configuration with respect to the planet azimuth. Figure~\ref{fig:B_field_model1} and \ref{fig:B_field_model2} show that the stagnation point is shifted in the azimuthal direction with respect to the planet location: in Model~1 the stagnation point has a negative azimuth, while it has a positive azimuth in Model~2. The sign of this asymmetry is the same in the MHD and the hydrodynamical cases, but the amplitude of the asymmetry is larger in the presence of the magnetic field. In Section~\ref{sec:torque}, we will show by varying the disc profiles that this is a general result. The origin of the stagnation point asymmetry in the hydrodynamic case will be further discussed in the next few paragraphs. For now, we note that the asymmetry in the stagnation point azimuth induces an asymmetry in the distribution of magnetic pressure. Indeed the magnetic field is more concentrated near the stagnation point, and the magnetic pressure is therefore maximum on the same side of the planet.

Finally, we compare the MHD simulations with a passive magnetic field (middle panels of Figures~\ref{fig:B_field_model1} and \ref{fig:B_field_model2}) to  a passive evolution 
found by solving  an advection/diffusion equation for the magnetic field
with a  fixed velocity field  (left panels) obtained from  a linear calculation
of the disc response to forcing by  the planet.
To perform these calculations the velocity field was constructed from
linear response calculations  for azimuthal mode numbers $m$ up to $60$  \citep[see][]{pp08}.
Singularities in the linear response were dealt with using the Landau prescription
which adds a purely imaginary frequency shift to the forcing frequency. The magnitude of the  shift was taken to be
$10^{-3}$ of the frequency itself, though as only a logarithmic singularity is involved in the worst case, changing this
to $10^{-5}$ was found to make no difference. This feature affects the solution only on the corotation circle 
where the streamlines turn. This  part of the flow 
 requires a nonlinear analysis,  as it is not  correctly represented
by linear theory \citep{pp08}. Nonetheless,
although the horseshoe region is somewhat narrower than that found from nonlinear simulations, 
as can be seen from  Figures~\ref{fig:B_field_model1} and \ref{fig:B_field_model2} the procedure reproduces the correct
asymmetry in the location of the stagnation point. 
 Such an asymmetry was already observed in hydrodynamical simulations by \citet{cm09} and \citet{pp09b}. 
\citet{pp09b} showed that it was due to an asymmetry between the inner and outer wakes. Indeed they showed that the azimuthal asymmetry of the stagnation point disappears  if the gravity of the planet is cut off at distances larger than H, 
 which prevents the excitation of the wake while keeping the horseshoe dynamics. 

In order to determine the kinematic effect on an initially purely azimuthal field, we solved the induction equation using the velocity field calculated with the linear response of the disc to the planet, and with the same magnetic diffusivity as in the full MHD simulations. The numerical domain is centred on the planet and extends over the full $2\pi$ in the azimuthal direction, and over a length $0.5r_p$ in the radial direction. Other parameters of the calculations were as in models 1 and 2. For the problem on hand the induction equation reduces to an advection diffusion equation for the flux function
that is readily soluble as  an initial value problem. At azimuthal boundaries we imposed periodic boundary conditions and at radial boundaries the magnetic flux was held fixed at its initial value, so conserving the total azimuthal magnetic flux. The ultimately steady state solutions 
illustrated in Figures~\ref{fig:B_field_model1} and \ref{fig:B_field_model2} show good qualitative and quantitative agreement
with the non linear simulations for weak fields in that the field gets concentrated along outgoing separatrices
with an azimuthal  asymmetry in the field strength that produces a larger
field on the side of the planet containing the stagnation point. 
The important feature is  the asymmetry of the stagnation point in  azimuth and the associated
asymmetry in field strength, which seems to be the main determinant of the corotation torque excess (see Section~\ref{sec:torque}). This  is quite well reproduced by  linear calculations and being  a linear feature, it should apply generally to the  low mass planet regime.

\section{An additional corotation torque of magnetic origin}
	\label{sec:torque}

\subsection{Two particular cases}
	\label{sec:torque_model1-2}

\begin{figure} 
  \centering
   \includegraphics[width=\columnwidth]{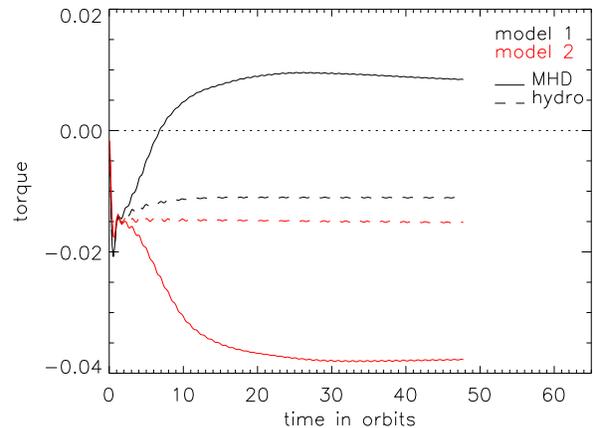}
  \caption{Time evolution of the torque exerted on the planet in models 1 and 2 (shown in black and red respectively), with and without a magnetic field (full and dashed lines respectively).}
  	\label{fig:torque_time_models1-2}
\end{figure}

\begin{figure*} 
  \centering
    \includegraphics[width=\columnwidth]{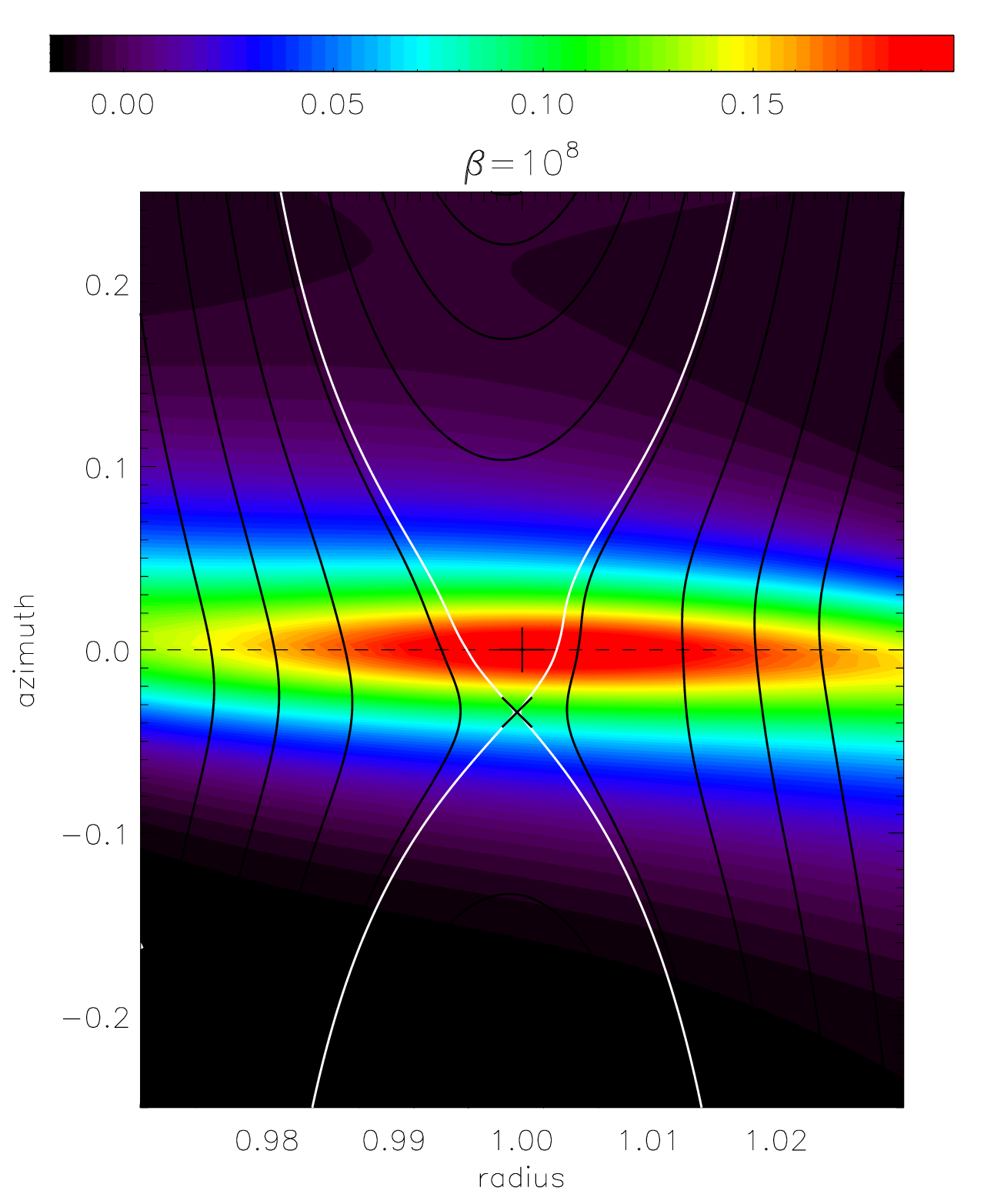}
    \includegraphics[width=\columnwidth]{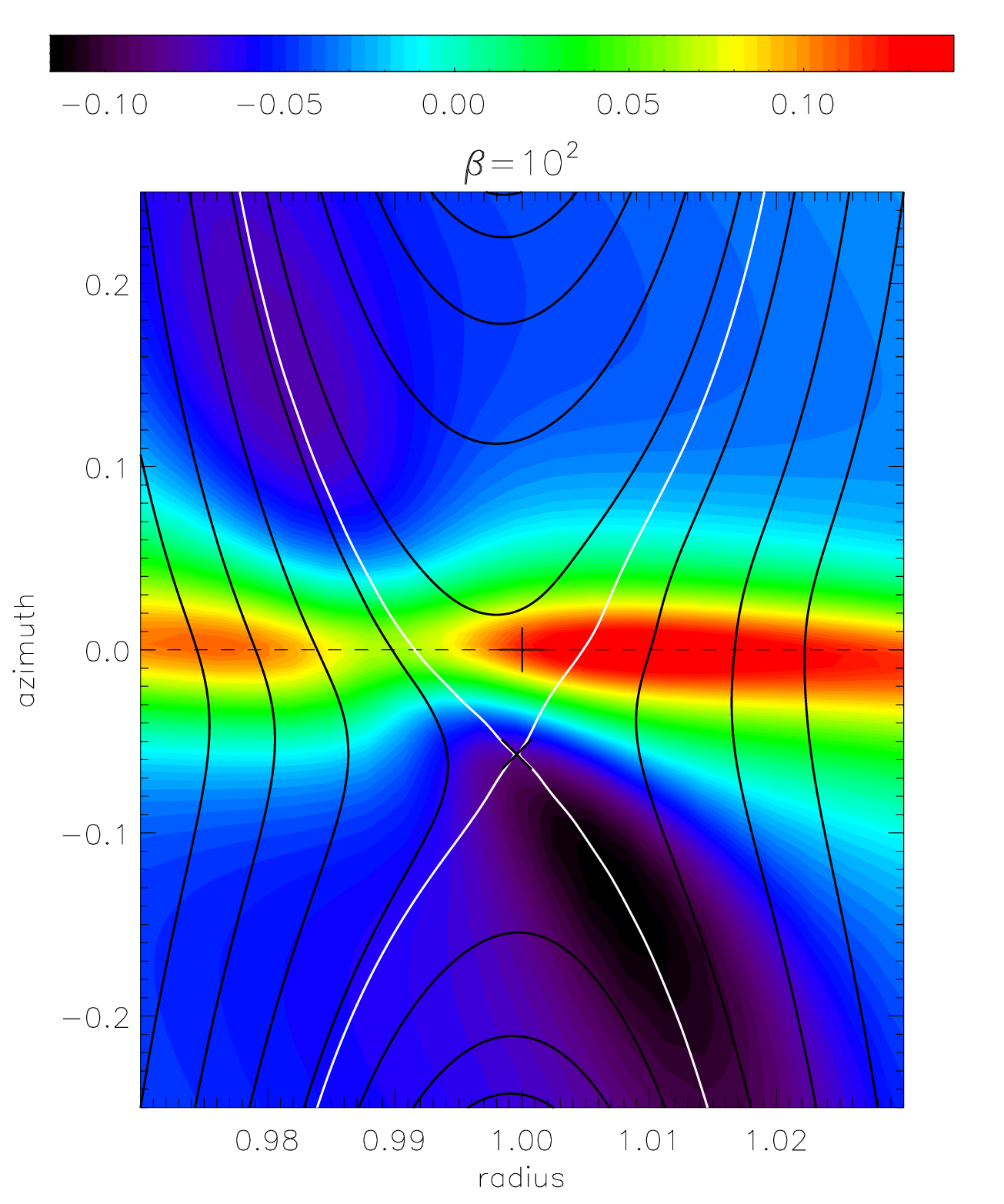}
        \includegraphics[width=\columnwidth]{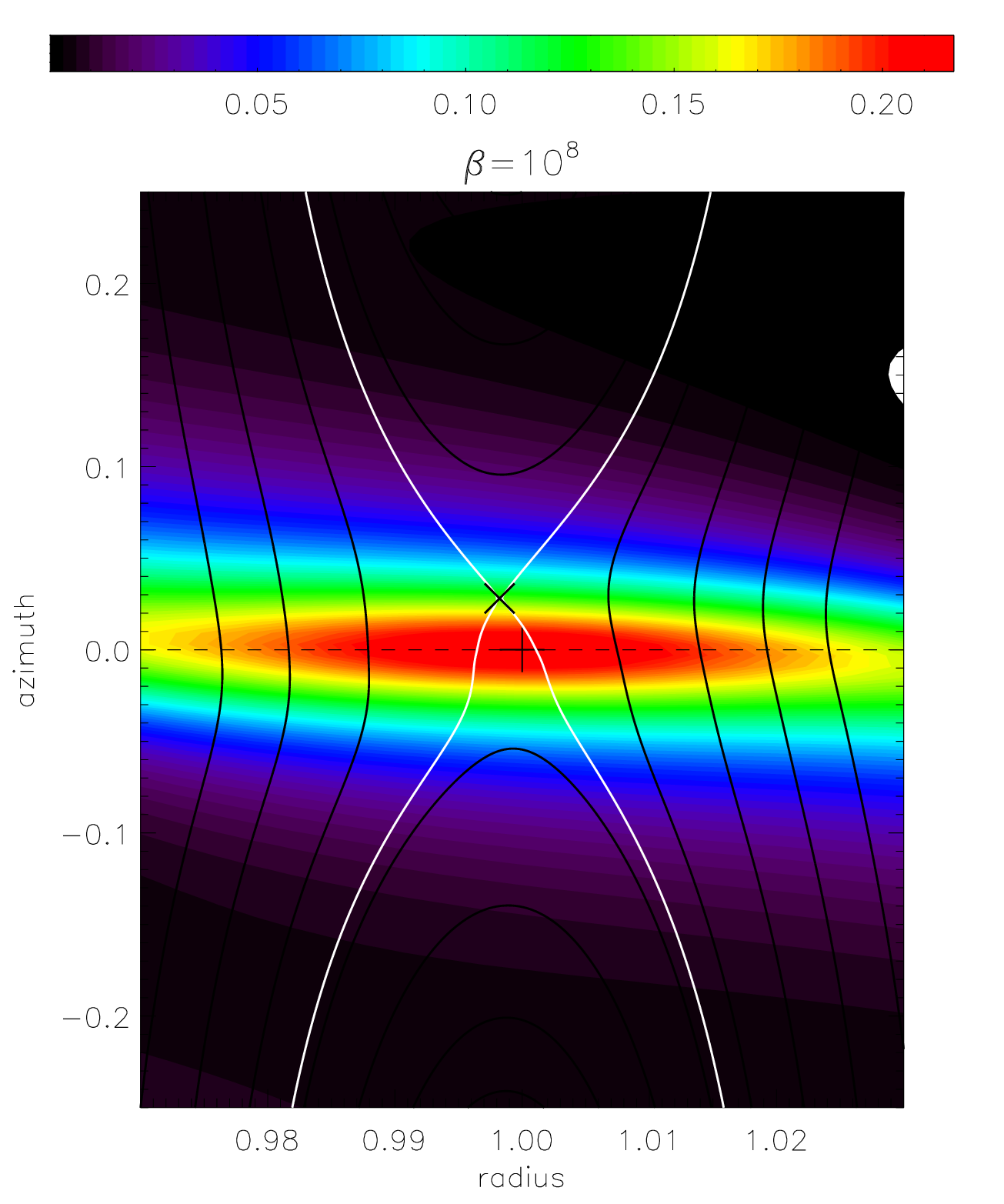}
    \includegraphics[width=\columnwidth]{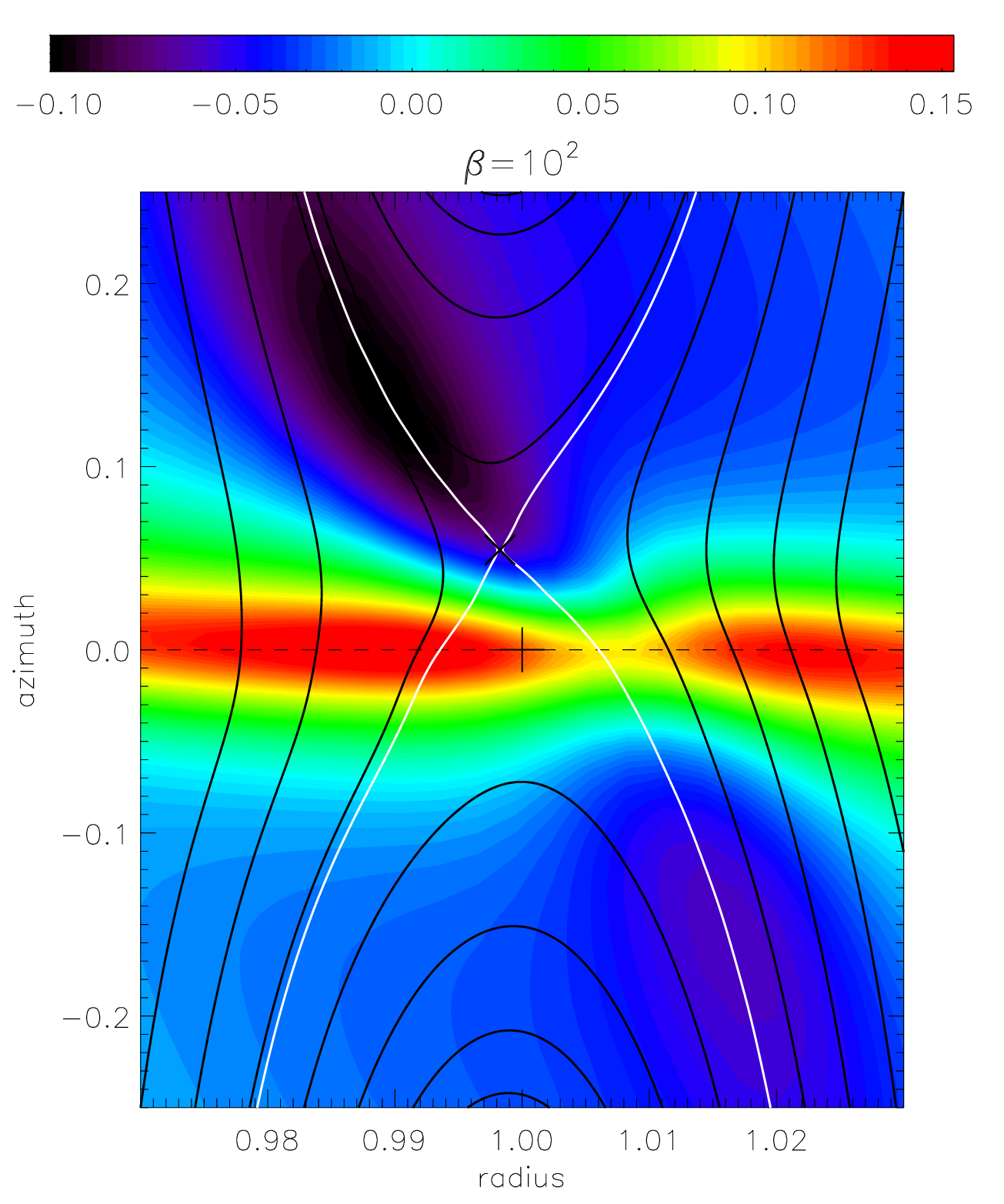}
  \caption{Comparison of the density perturbation (i.e. the density in the stationary state minus the density in the initial state) in hydrodynamical (left panels) and MHD simulations (right panels). Results are shown for model 1 (upper row), and model 2 (lower row). The black lines represent the streamlines, the white lines the separatrices of the horseshoe region, and the black cross the stagnation point.}
  \label{fig:drho}
\end{figure*}

We now describe the torque exerted by the disc on the planet in the MHD and corresponding hydrodynamical simulations, and start by considering the two fiducial models described in Section~\ref{sec:magnetic}. As shown in Figure~\ref{fig:torque_time_models1-2}, the torque is dramatically changed in the presence of a magnetic field. An additional torque component is observed in both Model~1 and 2, which takes about 10-15 orbits to establish. This corresponds to a few times the horseshoe U-turn timescale, which is about 3 orbits. This additional corotation torque has a large value with these fiducial parameters, and the total torque may become positive (in model~1). This underlines the potential importance of a weak magnetic field on type I migration, since here the magnetic pressure is only one percent of the thermal pressure. 

While the amplitude of the additional torque is very similar in both models, its sign differs: it is positive in Model~1 and negative in Model~2. To investigate this sign difference, we show the density perturbation induced by the planet for both models in Figure~\ref{fig:drho}. Comparing the hydrodynamical simulations (left panels) and MHD counterparts (right panels), we observe that the magnetic field induces underdense lobes located around the downstream separatrices. This can be interpreted as being due to the accumulation of magnetic pressure at the same location, which leads to a decrease of thermal pressure (and thus of surface density) to maintain approximate total pressure balance. We also notice a clear rear/front asymmetry in the density perturbation inside the planet's horseshoe region. The amplitude of the negative density perturbation takes its maximum value on the side of the planet where the stagnation point lies: at negative azimuth for Model~1, and positive azimuth for Model~2. This is consistent with the perturbation of magnetic pressure being maximum about the stagnation point. The origin and the sign of the additional torque can then be interpreted with this asymmetry of the density distribution. An underdense region induces a negative torque if it is in front of the planet (i.e. at positive azimuth), and a positive torque if it lies behind the planet (i.e. at negative azimuth). When the stagnation point has a positive azimuth as in Model~2, the largest underdensity lies in front of the planet and the additional torque is therefore negative. Conversely, when the stagnation point has a negative azimuth as in Model~1, the largest underdensity is behind the planet and the additional torque is therefore positive.  

Although the above interpretation of the MHD torque excess explains well the sign of the torque, we should add a note of caution in interpreting the torque using density lobes. Indeed, \citet{mc09} showed that such an interpretation could be misleading in the case of the corotation torque in an adiabatic hydrodynamical disc, because a diffuse density perturbation plays an important role as well. Instead they argued that the torque should be interpreted as arising from a creation of vortensity at the separatrices of the horseshoe region. In the case we are considering here, the magnetic field is also responsible for a creation of vortensity. This might be used to give an alternative explanation of the torque excess, although it is not yet clear whether a similar interpretation applies in the presence of a magnetic field.

\subsection{Dependence on the gradients of density, temperature and magnetic field strength}
	\label{sec:torque_gradients}
We have shown in the previous section that a new component of the corotation torque arises in the presence of a weak magnetic field. Its amplitude is the same but its sign differs in two disc models that differ by the choice of the temperature and density profiles. To better characterise this new torque component, we performed series of MHD and HD runs in which we varied the slopes of the background magnetic field, density and temperature profiles. The results of these series of run are presented in the next subsections.

\subsubsection{Varying the magnetic field strength gradient}
\label{sec:torque_b}	
\begin{figure} 
  \centering
   \includegraphics[width=\columnwidth]{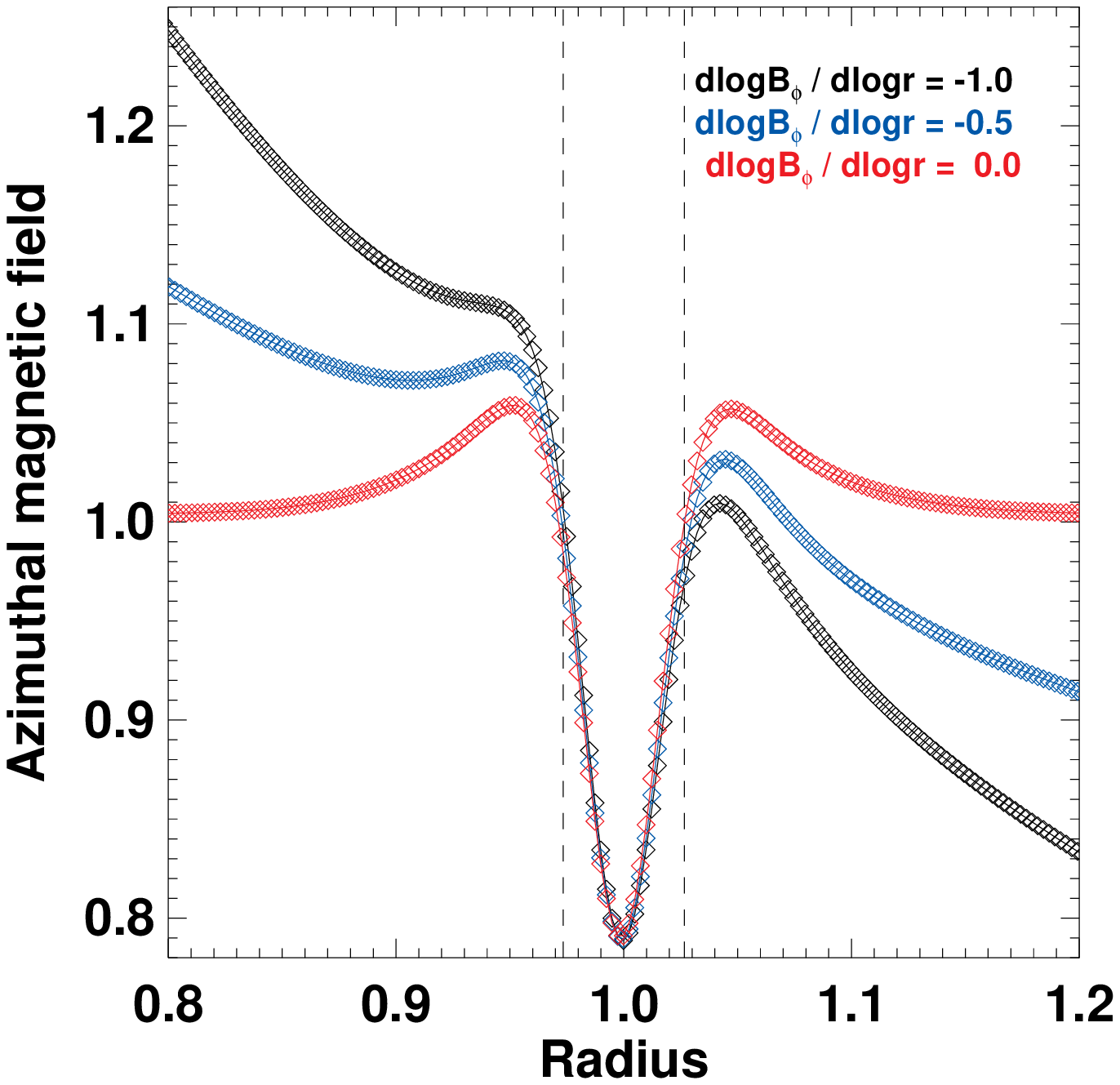}
   \includegraphics[width=\columnwidth]{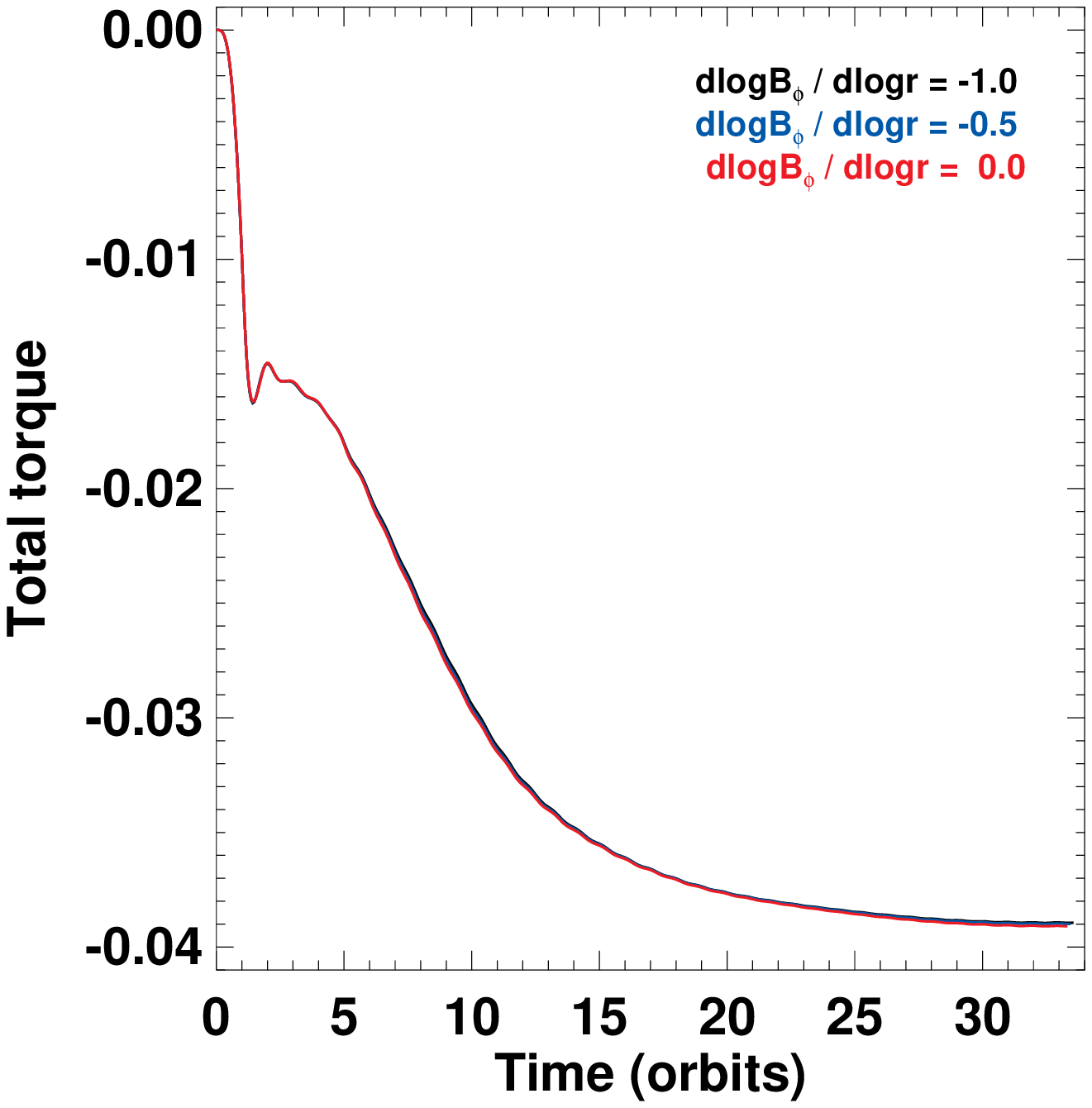}
  \caption{Results of simulations with different slopes of the initial magnetic field profile : $b=-1$ (black), $b=-0.5$ (blue) and $b=0$ (red).  Upper panel : azimuthally averaged magnetic field profile in a steady state normalised by its initial value at the planet location (the vertical dashed lines show the extent of the horseshoe region). Lower panel : time evolution of the torque. Strikingly the torques from the different simulations are indistinguishable.}
  	\label{fig:B_profiles}
\end{figure}

Strikingly, the torque does not depend at all on the gradient of magnetic field strength ! This is illustrated in Figure~\ref{fig:B_profiles}, showing the azimuthally averaged magnetic field strength (upper panel) and the time evolution of the total torque (lower panel). Three models with different slope of the magnetic field strength are shown : $b=-1,-0.5$ and $0$. The magnetic field profiles differ substantially (except within the planet's horseshoe region) but the time evolution of the torque in the different models is indistinguishable.

This surprising result contrasts with the usual behaviour of the corotation torque, which is often proportional to the gradient of some conserved quantity (vortensity in an isothermal disc, entropy for a disc with an energy equation). It also differs from the regime of strong magnetic field where the torque arises from magnetic resonances :  \citet{terquem03} indeed found that the sign and magnitude of the torque depends strongly on the slope of the magnetic field strength profile. How should this different behaviour be interpreted ? One might interpret the independence on the magnetic field gradient by the fact that the horseshoe motion redistribute the magnetic flux in the horseshoe region (it tends to accumulate it on the downstream separatrix). Therefore what matters is the total flux in the horseshoe region rather how it is initially  distributed in it. The usual proportionality to gradients is due to the fact that gradients are the cause of a symmetry breaking, necessary for the presence of a net torque. Here the presence of a magnetic field induces an azimuthal asymmetry of the horseshoe motion and of the density distribution, which is not proportional to any gradient. Indeed, it will be shown later that the sign of this asymmetry is governed by the gradients of density and temperature, but its amplitude depends only on the magnetic field strength. Thus the MHD torque excess is not proportional to any gradient because an asymmetry seems to be caused by the interaction between the magnetic field and the horseshoe motion.

\subsubsection{Varying the temperature gradient}
	\label{sec:torque_q}
\begin{figure} 
  \centering
    \includegraphics[width=\columnwidth]{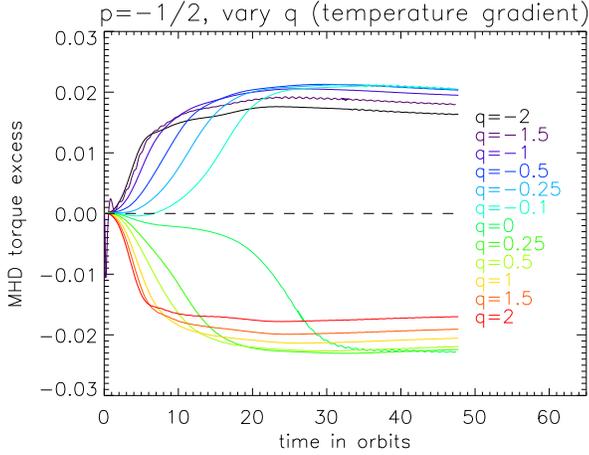}
  \caption{Time variation of the MHD torque excess (defined as the torque in the MHD simulation minus the torque in its hydrodynamical counterpart) for various values of the temperature slope index $q$, and a density profile  $p=-1/2$.}
      \label{fig:torque_time_q_diff}
\end{figure}

\begin{figure} 
  \centering
    \includegraphics[width=\columnwidth]{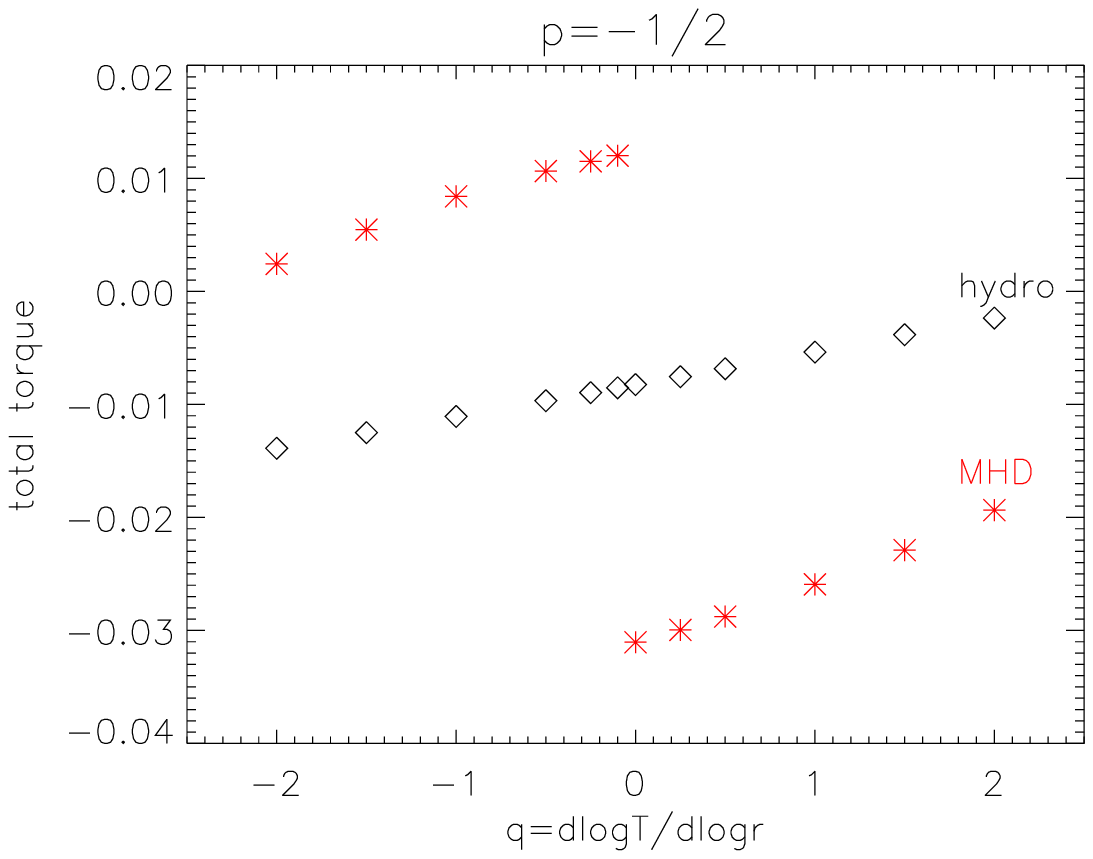}
     \includegraphics[width=\columnwidth]{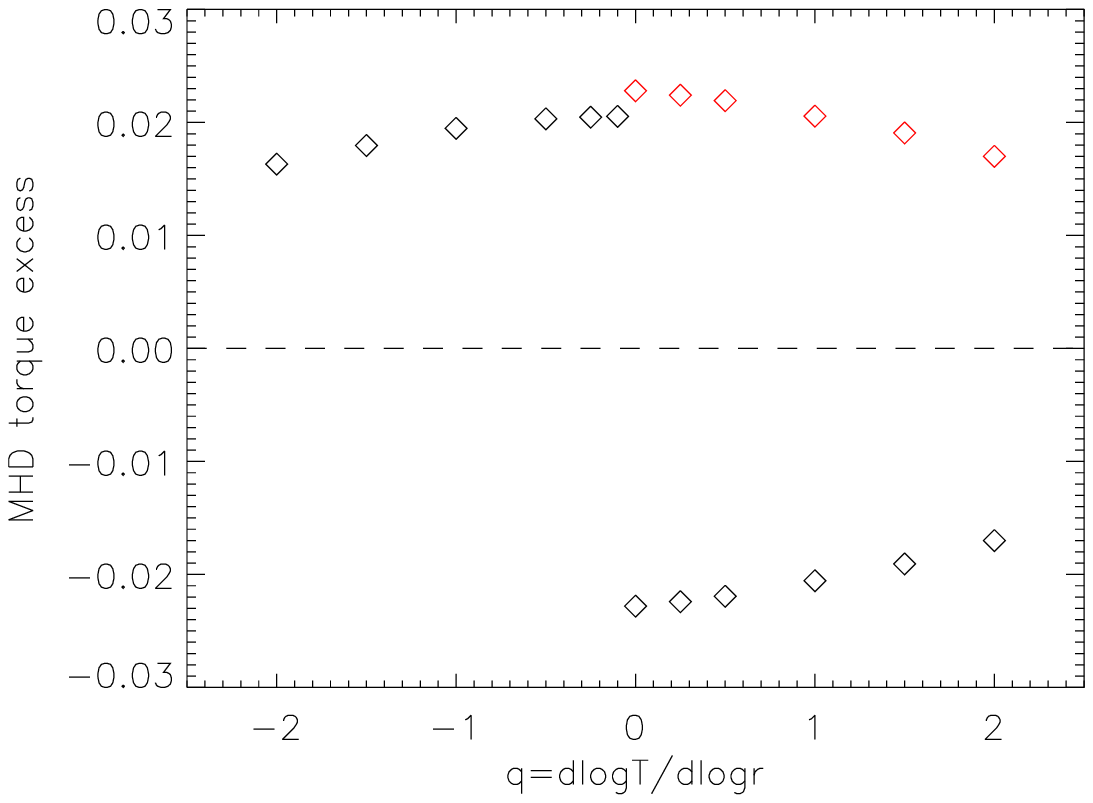}
    \includegraphics[width=\columnwidth]{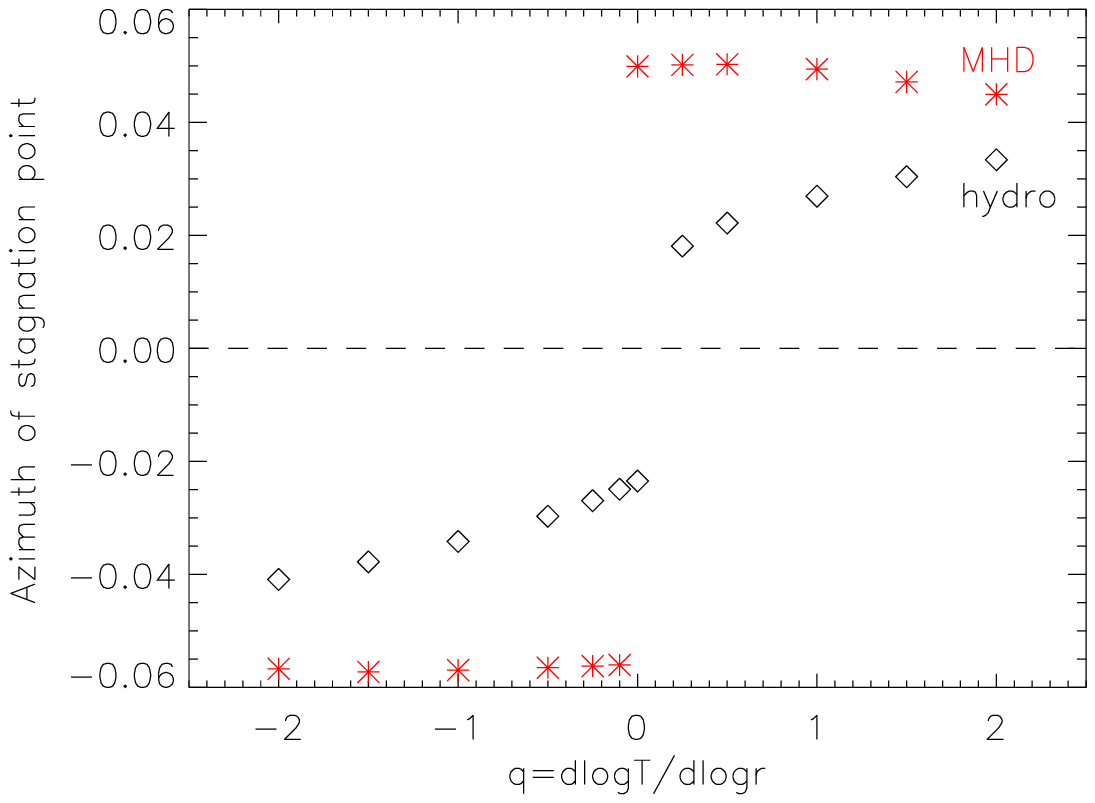}
  \caption{Results of a series of simulations varying $q$ for $p = -1/2$. Upper panel: total torque in a steady-state as a function of the slope of the background temperature profile, for MHD (red stars) and HD simulations (black diamonds). Middle panel: MHD torque excess, defined as the MHD torque subtracted by the hydrodynamical torque. Red star symbols show the absolute value of the torque excess (when it is negative). Lower panel: Azimuth of the stagnation point in hydrodynamical (black) and MHD simulations (red). The dependence of $\phi_s$ with the temperature gradient is similar to that of the outer X stagnation point obtained in Fig. 10 of \citet{cm09} (the other two stagnation points appear only at smaller softening length, as discussed in Section~\ref{sec:softening}).}
  	\label{fig:p-1/2_vary_q}
\end{figure}

\begin{figure} 
  \centering
    \includegraphics[width=\columnwidth]{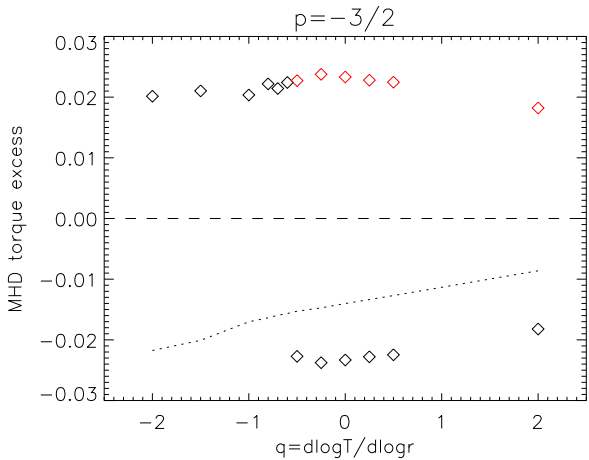}
  \caption{Same as the middle panel of Figure~\ref{fig:p-1/2_vary_q} but with $p=-3/2$: MHD torque excess (defined as the MHD torque subtracted by the hydrodynamical torque) as a function of the temperature slope index $q$. Red star symbols show the absolute value of the MHD torque excess (when it is negative). For comparison the dotted line shows the hydrodynamical torque. }
    	\label{fig:p-3/2_vary_q}
\end{figure}

Figure~\ref{fig:torque_time_q_diff} shows the time evolution of the MHD torque excess (defined as the torque in the MHD simulation minus the torque in its hydrodynamical counterpart) for a series of runs varying the temperature gradient. It shows that the torque excess changes sign abruptly around $q=0$, while its amplitude in steady state remains roughly independent of $q$ within 10-20\%. This behaviour is also displayed on the top and middle panels of Figure~\ref{fig:p-1/2_vary_q} showing respectively the torque and the MHD torque excess as a function of the index of the temperature profile $q$. Importantly, this discontinuous behaviour of the torque excess follows that of the azimuth of the stagnation point, shown in the lower panel of Figure~\ref{fig:p-1/2_vary_q}. Indeed the azimuth of the stagnation point changes sign abruptly around the same temperature gradient as the torque excess ($q=0$), in the MHD as well as the hydrodynamical simulations. This confirms the interpretation of the MHD torque excess given in Section~\ref{sec:torque_model1-2} as an asymmetry of the density distribution in the horseshoe region driven by the azimuthal shift of the stagnation point with respect to the planet. It also confirms that the sign of the stagnation point azimuth is the same in the MHD and the hydrodynamical cases, as was noted in the particular cases of Model~1 and 2. The amplitude of the azimuthal shift is however different: in the MHD case it appears to be independent of the temperature gradient, while in the hydrodynamical case it decreases when approaching the transition where it changes sign. Thus the sign of the stagnation point azimuth is governed by a hydrodynamical process determined by the density and temperature gradients (linked with an asymmetry between the inner and outer wakes as discussed in Section~\ref{sec:magnetic}), while its amplitude seems to be driven by a magnetic process independent of these gradients (its dependence on the magnetic field strength will be discussed in Section~\ref{sec:alpha_beta}). Finally, we note that while the final amplitude of the torque barely depends on the temperature gradient, the time evolution to reach this amplitude differs significantly: the closer to the transition where the torque excess changes sign abruptly, the longer it takes for the torque excess to build up (Figure~\ref{fig:torque_time_q_diff}). This may be interpreted in the following manner. The presence of the magnetic field tends to amplify any azimuthal asymmetry present in the hydrodynamical flow to finally reach an amplitude independent of the initial asymmetry. It may therefore seem natural that a smaller initial asymmetry takes longer to be amplified by the magnetic field, resulting in the time evolution of the torque excess observed in Figure~\ref{fig:torque_time_q_diff}.

Figure~\ref{fig:p-3/2_vary_q} shows the dependence of the MHD torque excess with the temperature gradient using a different density profile ($p=-3/2$). This confirms the qualitative features observed with $p=-1/2$: the amplitude of the torque excess is roughly constant, but its sign changes abruptly at a critical temperature gradient. The value of the temperature gradient at the transition depends on the density gradient (it is $q\simeq-0.55$ for $p=-3/2$, and $q\simeq-0.05$ for $p=-1/2$) but it still corresponds to the transition where the stagnation point azimuth changes sign in both hydrodynamical and MHD simulations (not shown here). The dependence on the density profile is further investigated in the following subsection.

\subsubsection{Varying the density gradient}
	\label{sec:torque_p}
	
\begin{figure} 
  \centering
      \includegraphics[width=\columnwidth]{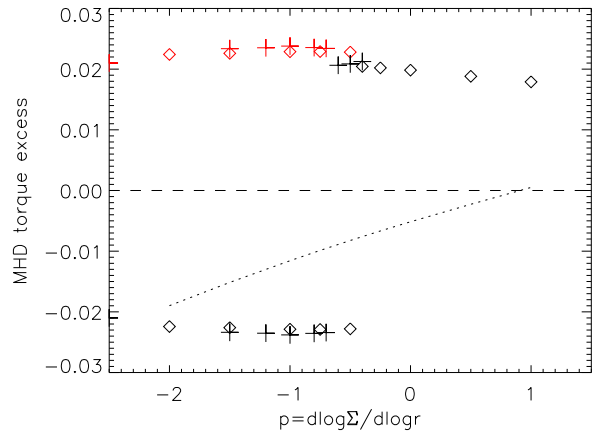}
  \caption{Results of a series of runs where the density profile is varied at fixed uniform temperature profile ($q=0$).  The MHD torque excess is shown as a function of the slope index of the density profile $p$. Red symbols show the absolute value of the torque excess. For comparison the dotted line shows the hydrodynamical torque. The simulations were performed using both the codes RAMSES (diamonds) and NIRVANA (plus signs), which allows to assess the numerical errors. The two codes agree within $5\%$ for the amplitude of the torque excess, and find the sign transition at $q=-0.6\pm0.2$. }
  	  \label{fig:q0_vary_p}	
\end{figure}

We performed a series of runs varying the density profile at fixed uniform temperature profile ($q=0$). Figure~\ref{fig:q0_vary_p} shows the dependence of the MHD torque excess with the slope index of the density profile $p$. The behaviour is very similar to that described in Section~\ref{sec:torque_q} when varying the temperature profile: the amplitude of the torque excess is barely dependent on the density profile, but its sign does depend on it. A sharp transition from a negative to a positive MHD torque excess is obtained at $p=-0.55\pm0.15$ (between $-0.7$ and $-0.6$ for NIRVANA, and between $0.5$ and $0.4$ for RAMSES). Therefore, steeply decreasing density profiles give a negative MHD torque excess, while shallower (or increasing) profiles drive a positive MHD torque excess.

Figure~\ref{fig:q0_vary_p} can also be used to assess the importance of numerical errors, since the series of simulations has been performed both with the RAMSES and NIRVANA codes. We find that the qualitative behaviour of constant amplitude and sharp sign transition is a robust feature of the simulations. The amplitude of the torque excess agrees within $5\%$ between the two codes, and the location of the sign transition agrees within $\Delta p = 0.2$.

\subsubsection{Sign of the MHD torque excess}

\begin{figure} 
  \centering

    \includegraphics[width=\columnwidth]{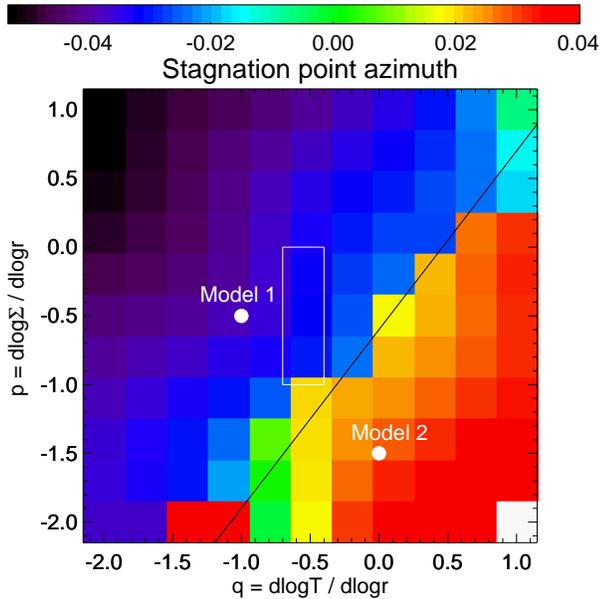}
  \caption{Azimuth of the (outer-X) stagnation point ($\varphi_s$) in a series of non-magnetised simulations, as a function of the background surface density and temperature gradients at the planet orbit. The solid line shows the linear function $p-1.3q+0.6=0$. Wherever $\varphi_s < 0$, the magnetic-related horseshoe drag is positive, while it is negative for $\varphi_s > 0$. The position of the two fiducial models is represented with a white circle. The white rectangle represents the range of parameters suggested by observations of protoplanetary discs at large radii: $q \in [-0.7,-0.4]$, $p\in[-1,0]$ (see Section~\ref{sec:discussion} for a discussion).}
  	\label{fig:phistagn_hydro}
\end{figure}

We have found that the amplitude of the MHD torque excess is roughly independent of the temperature, density and magnetic field strength gradients. On the other hand, the sign of the torque excess does depend on both the temperature and density gradients. This sign is dictated by the sign of the stagnation point azimuth (which is the same in both hydrodynamical and MHD simulations): a positive stagnation point azimuth leads to a negative torque excess, while a negative azimuth leads to a positive torque excess. This is interpreted as due to a rear-front asymmetry of the underdense lobes near the downstream separatrices, the strongest underdensity arising on the side of the planet where the stagnation point lies.

Given that the azimuth of the stagnation point is observed to have the same sign in HD and MHD simulations, 
one can use HD simulations to infer its sign and thus the sign of the MHD torque excess.
 For this purpose, we have run HD simulations varying both the density and temperature profiles.
 The simulations were performed with FARGO: a two-dimensional hydrodynamical code similar to NIRVANA,
 which uses the so-called Fargo algorithm to speed up the calculation by increasing the time-step \citep{fargo1}.
 The results are presented in Figure~\ref{fig:phistagn_hydro} showing the azimuth of the stagnation in the plane $\{p,q\}$. A region of positive azimuth at low $p$ and large $q$ is clearly separated from a region where the azimuth is negative at large $p$ and low $q$. The limit between these two regions of parameter space can be approximated by a linear function of $p$ and $q$: $p-1.3q+0.6=0$. Thus we find that the sign of the stagnation point azimuth is given by:
\begin{equation}
{\rm sign}(\varphi_s) = {\rm sign}(1.3q - p - 0.6).
	\label{eq:phistagn_sign}
\end{equation}
As mentioned before, the MHD torque excess has the opposite sign. This approximate formula can now be compared to the results of Section~\ref{sec:torque_model1-2}, \ref{sec:torque_q} and \ref{sec:torque_p}. We first note that Model~1 (respectively 2) lies in the region of negative (positive) stagnation point azimuth, in agreement with the results of Section~\ref{sec:torque_model1-2}. At $p=-0.5$, the formula predicts a sign transition at $q=0.08$, to be compared with $q\simeq-0.05$ found in Section~\ref{sec:torque_q} (Figure~\ref{fig:p-1/2_vary_q}). At $p=-3/2$, the formula gives $q=-0.7$ against $q\simeq-0.55$ for the simulations of Figure~\ref{fig:p-3/2_vary_q}. Finally, at $q=0$ the formula predicts a sign transition at $p=-0.6$ against $p=-0.55\pm0.15$ for the simulations of Figure~\ref{fig:q0_vary_p}. All in all, equation~\ref{eq:phistagn_sign} therefore reproduces the results of the simulations with a precision of $0.1-0.2$ in $p$ and $q$.

\subsection{Dependence on magnetic field strength and diffusion coefficients}
	\label{sec:alpha_beta}
\begin{figure} 
  \centering
     \includegraphics[width=\columnwidth]{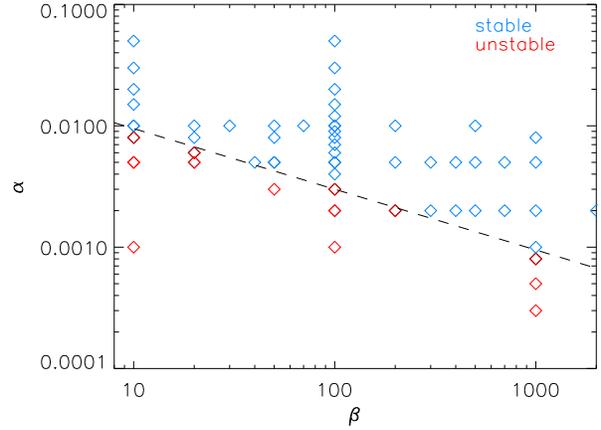}
  \caption{Diagram showing stable (blue diamonds) and unstable (red diamonds) runs in the $\{\alpha,\beta\}$ plane. The dashed line corresponds to a fit of the marginal stability: $\alpha\sqrt{\beta} = 3.10^{-2}$.}
  	\label{fig:stability}
\end{figure}

\begin{figure} 
  \centering
    \includegraphics[width=\columnwidth]{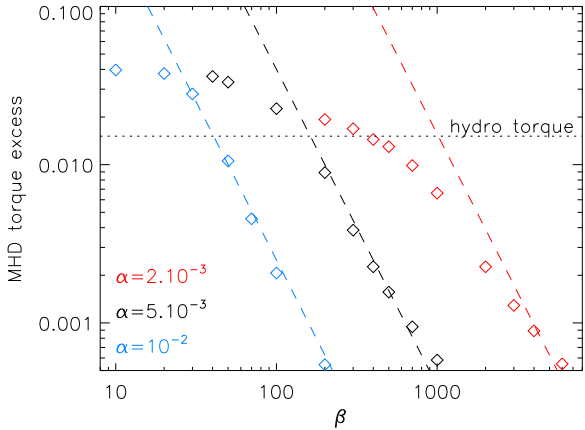}
  \caption{Amplitude of the MHD torque excess as a function of $\beta$ for three different values of $\alpha$: $2.10^{-3}$ (red), $5.10^{-3}$ (black) and $10^{-2}$ (green). The dashed lines show the fit given by Equation~\ref{eq:fit} (which assumes $ \Gamma \propto 1/(\beta^2\alpha^4)$). For strong magnetic fields or low diffusion coefficients the torque excess deviates from this scaling. Other parameters are the fiducial ones and the disc Model~2.}
  	\label{fig:torque_beta}
\end{figure}

\begin{figure} 
  \centering
    \includegraphics[width=\columnwidth]{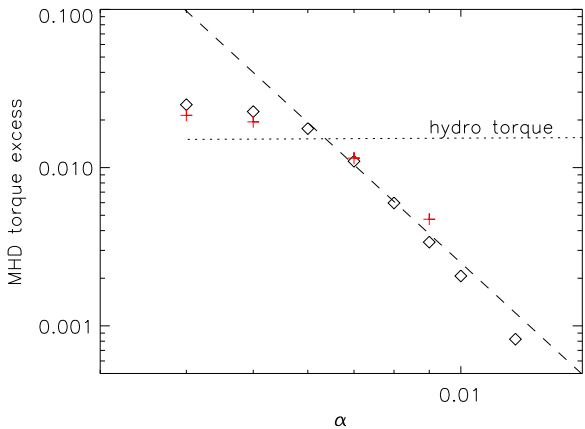}
  \caption{Amplitude of the MHD torque excess as a function of $\alpha$ for $\beta=100$. Other parameters are the fiducial ones and the disc Model~2 (black diamonds) or Model~1 (red $+$ signs). Here, the dimensionless resistivity $\alpha/\Pm$ (which we show in Figure~\ref{fig:torque_Pm} to be the relevant diffusion parameter determining the MHD torque excess) is equal to $\alpha$ with our fiducial value of $\Pm=1$. The dashed line shows the fit given by Equation~\ref{eq:fit} with $ \Gamma \propto (\Pm/\alpha)^4$. For low values of $\alpha$ the torque excess deviates from this scaling. }
  	\label{fig:torque_alpha}
\end{figure}

\begin{figure} 
  \centering
    \includegraphics[width=\columnwidth]{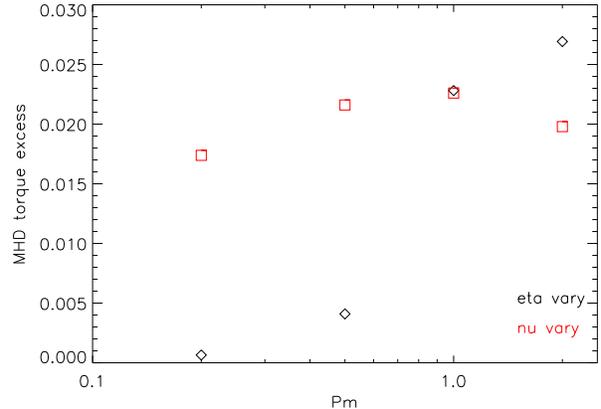}
  \caption{Amplitude of the MHD torque excess as a function of the magnetic Prandtl number. The black diamonds show a series of run where the viscosity is kept constant to $\alpha=5.10^{-3}$, while the resistivity is varied. Reversely, the red squares correspond to the resistivity being constant $\alpha/\Pm=5.10^{-3}$, while the viscosity is varied. The magnetic torque excess has a steep dependence on the resistivity, but only a very weak one on the viscosity (for the range of diffusion coefficients considered). Other parameters are the fiducial ones and the disc Model~2.}
  	\label{fig:torque_Pm}
\end{figure}

As shown in Section~\ref{sec:torque_gradients}, the amplitude of the MHD torque excess does not depend on the gradients of magnetic field strength, density and temperature. It may therefore depend only on the magnetic field strength (through $\beta$), the diffusion coefficients ($\alpha$ and $\Pm$), the mass of the planet ($M_p/M_*$), the aspect ratio of the disc ($h$) and the softening length of the planet potential. In this section we explore the dependence on the magnetic field strength and the diffusion coefficients, and leave for future work the study of different values of the planet mass and aspect ratio of the disc. The effect of varying the softening length is briefly described in Section~\ref{sec:softening}

\subsubsection{Stability domain}
When varying $\alpha$ and $\beta$, we observed that for small values of these parameters the horseshoe motion could become unstable. This instability grows on the downstream separatrices of the horseshoe region. It is therefore due to the presence of the planet rather than an intrinsic property of the disc. As an additional check, we ran simulations without a planet but with small velocity perturbations added to the background flow, and found no instability. The instability is probably caused by a creation of vortensity due to the interaction of the magnetic field with the horseshoe motion. When the vortensity perturbation is strong enough, this leads to a shear instability that creates vortices near the downstream separatrices. A detailed description of this instability is however postponed to future work. Here we focus on the stable region of parameter space to characterise the dependence of the torque excess on both $\alpha$ and $\beta$.

The region of parameter space $\{\alpha,\beta\}$ where the horseshoe motions are stable or unstable is shown in Figure~\ref{fig:stability}. From this figure, we deduce the following criterion for the motion to be stable for our fiducial values of planet mass and aspect ratio:
\begin{equation}
\alpha\sqrt{\beta} > 3\times10^{-2}.
	\label{eq:stability}
\end{equation}
In the following Sections, we present the results of simulation in this parameter domain.

\subsubsection{MHD torque excess}
Figure~\ref{fig:torque_beta} shows the amplitude of the MHD torque excess in three series of runs varying $\beta$ for three different values of $\alpha$. As expected for a magnetic effect, the MHD torque excess increases with the magnetic field strength (i.e. when $\beta$ decreases). It also increases very steeply when the diffusion is decreased, as can be seen by comparing the different series in Figure~\ref{fig:torque_beta} and more directly in Figure~\ref{fig:torque_alpha} showing the dependence of the MHD torque excess on $\alpha$ for $\beta=100$. The latter also shows that the MHD torque excess vanishes for high values of the diffusion coefficients, as can be expected because the magnetic field and gas motion decouple when the resistivity is very large. The dependence on the diffusion can be interpreted in the following way. The MHD torque excess is due to an azimuthally shifted accumulation of magnetic pressure along the downstream separatrix. The efficiency with which the magnetic flux can be accumulated at the separatrix is governed by an equilibrium between the advection by the horseshoe motion and the diffusion by the resistivity: the lower the resistivity the thinner the layer where the magnetic flux is accumulated and therefore the stronger the magnetic pressure there. This stronger magnetic pressure leads to a stronger underdensity, thus explaining the steep increase of the MHD torque excess when the diffusion is decreased.

Following this reasoning, we expect that the steep dependence of the torque excess is mostly due to the resistivity rather than the viscosity (both of which are varied concurrently in these runs where the magnetic Prandtl number is fixed to $\Pm=1$). We checked this by performing two series of runs where one of the diffusion coefficients is varied while keeping the other one constant and show the results in Figure~\ref{fig:torque_Pm} as a function of the magnetic Prandtl number. Varying the viscosity by a factor 10 changes the MHD torque excess by no more than $20\%$, if the resistivity is kept constant. By contrast, a similar variation of the resistivity at fixed viscosity changes the torque excess by a factor 40 ! To first order, the MHD torque can therefore be considered to be independent of the viscosity in the range of parameters explored in this study. This is probably due to the fact that the values of the viscosity considered are close to that leading to a fully unsaturated horseshoe drag, around which the dependence on the viscosity is quite weak. It is expected that decreasing the viscosity further would eventually lead to a saturation of the corotation torque as in the hydrodynamical situation, due to a lack of angular momentum exchange between the coorbital region and the rest of the disc.

Figure~\ref{fig:torque_beta} suggests that the MHD torque excess has a power law dependence with respect to $\beta$ for moderately large values of $\beta$ and $\alpha$. We find that the torque excess is fairly well fitted with the following scaling (for our fiducial values of the planet mass and disc aspect ratio): 
\begin{equation}
\Gamma = 4.10^{-2} \left(\frac{100}{\beta}\right)^2 \left(\frac{5.10^{-3}\Pm}{\alpha}\right)^4.
	\label{eq:fit}
\end{equation}
This fit is overplotted in Figures~\ref{fig:torque_beta} and \ref{fig:torque_alpha} with dashed lines. It is valid for large enough values of $\alpha$ and $\beta$ corresponding to a MHD torque excess smaller or comparable to the torque in a hydrodynamical case. For smaller values of these two parameters, the dependence flattens out leading to a weaker torque excess than predicted by the above scaling. It is interesting to note that this scaling depends on the same combination of powers of $\alpha$ and $\beta$ as in the stability criterion (equation~\ref{eq:stability}): the value of the dimensionless parameter $\beta\alpha^2$ determines both the stability of the horseshoe motion and the amplitude of the MHD torque excess. Although the physical origin of this scaling remains somewhat uncertain and should be investigated further in the future, we propose the following tentative interpretation. The density perturbation in the horseshoe region can be expected to be inversely proportional to $\beta$ as it is proportional to the magnetic pressure. The torque excess scaling as $1/\beta^2$ may then be understood if the rear-front asymmetry of the density lobes is also inversely proportional to $\beta$ (see below the dependence of the stagnation point azimuth with $\beta$). One may further speculate that the density perturbation and rear-front asymmetry scale as $1/\alpha^2$, leading to the desired scaling of the torque excess. 

\subsubsection{Horseshoe region}

\begin{figure} 
  \centering
    \includegraphics[width=\columnwidth]{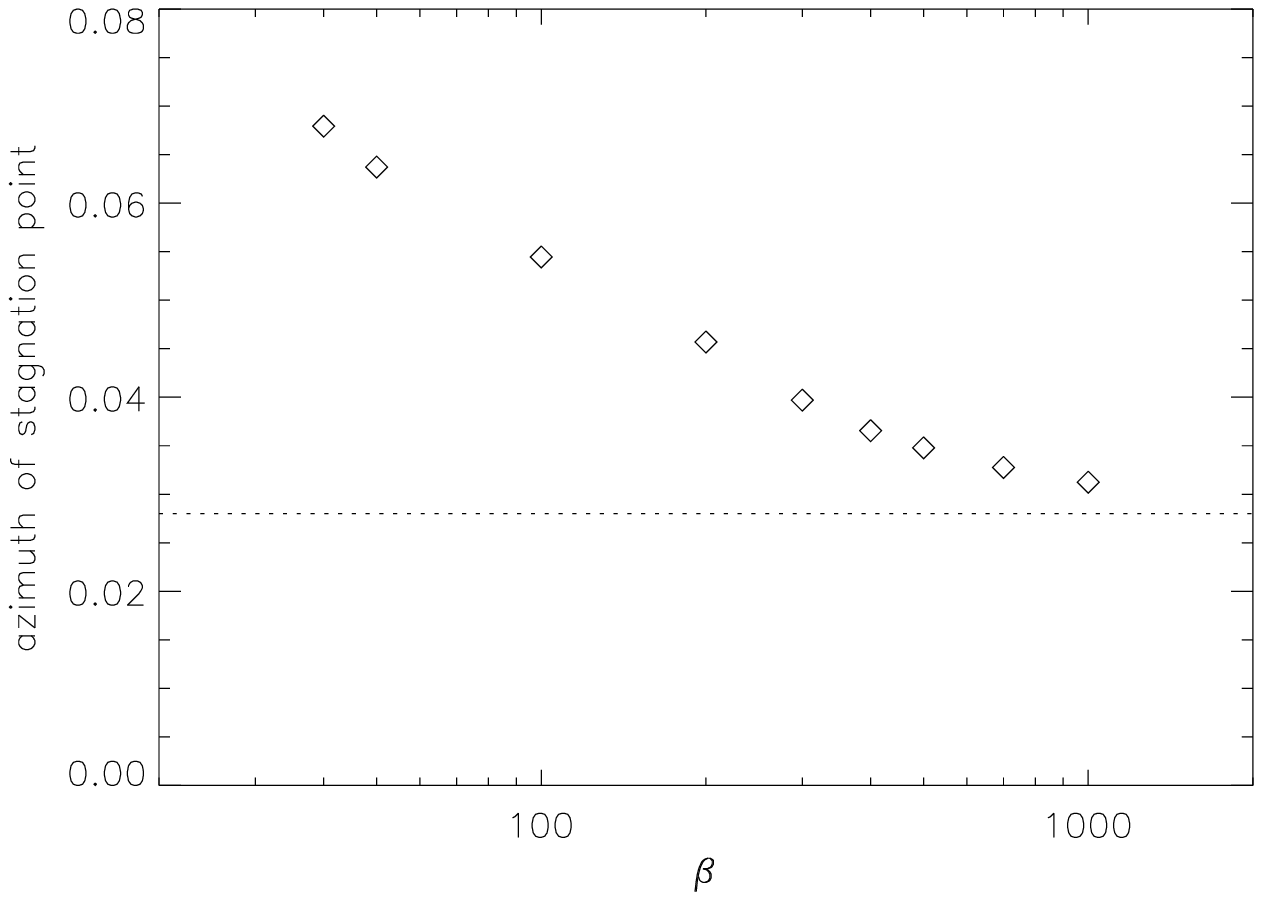}
    \includegraphics[width=\columnwidth]{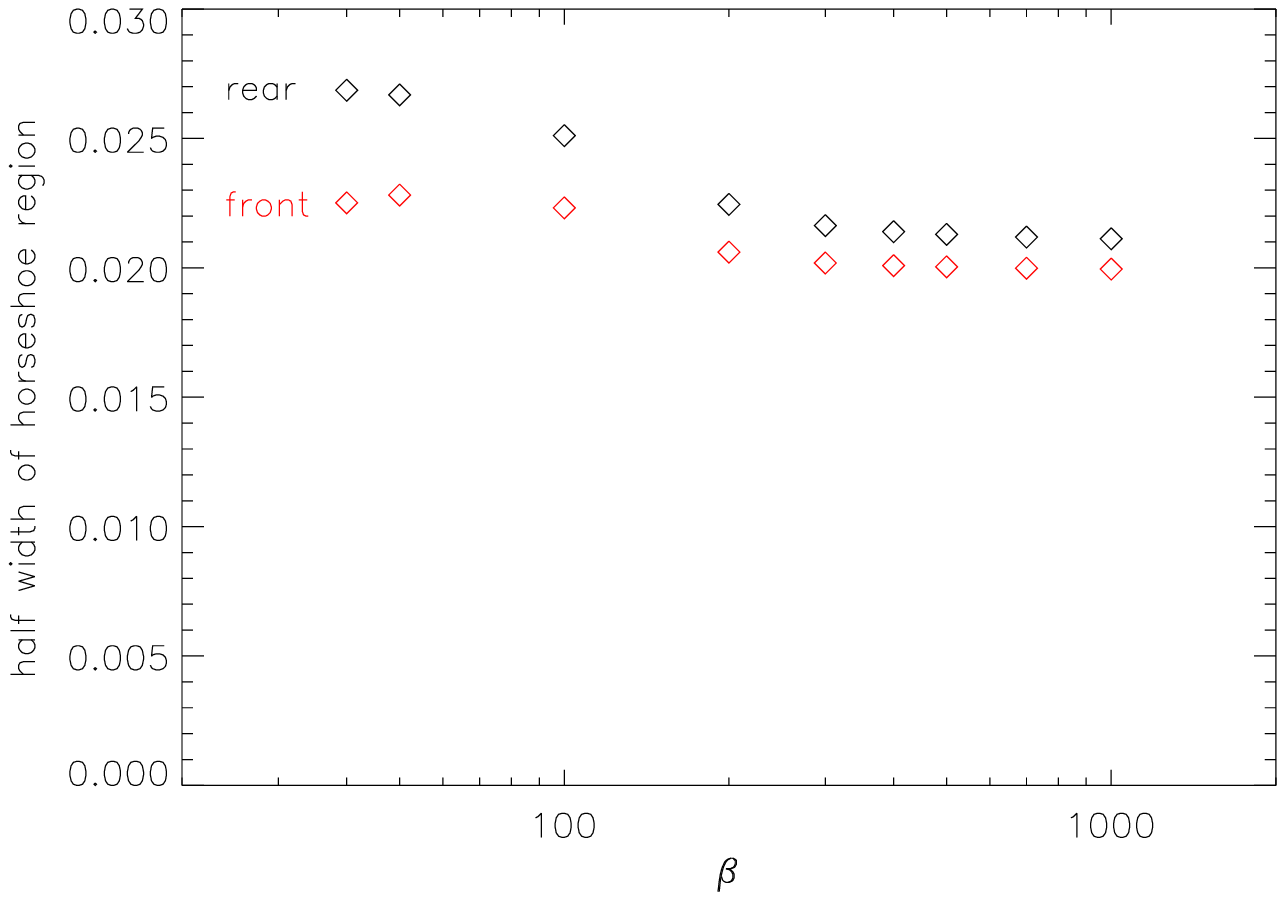}    
  \caption{Characteristics of horseshoe region as a function of $\beta$ for $\alpha=5.10^{-3}$ and model~2. Upper panel: azimuth of stagnation point. The dotted line shows the value for a non magnetised run. Lower panel: half width of horseshoe region measured at an azimuth of $\varphi=\pm1$. The rear ($\varphi=-1$) is shown in black, while the front ($\varphi=1$) is shown in red. Other parameters are the fiducial ones and the disc Model~2.}
  	\label{fig:phistagn_beta}
\end{figure}

After studying the torque, we turn to the characteristics of the horseshoe region and their dependence on the magnetic field strength. Figure~\ref{fig:phistagn_beta} shows the azimuth of the stagnation point and the half-width of the horseshoe region as a function of $\beta$. At large $\beta$, the azimuth of the stagnation point and the half width of the horseshoe region tend to their hydrodynamical value, as expected in the weak magnetic field limit. As the magnetic field increases, the azimuth of the stagnation point increases significantly, confirming the idea that the presence of the magnetic field amplifies the azimuthal asymmetry of the horseshoe motion. The width of the horseshoe region also increases with stronger magnetic fields (in this weak field regime). This result is somewhat surprising because one may naively expect the magnetic tension to resist the horseshoe motion, and therefore that the width of the horseshoe region decreases with stronger magnetic field. The reason for the observed behaviour remains unclear. We also observe that the rear-front asymmetry of the horseshoe width is more and more pronounced as the magnetic field is increased.

\subsection{Dependence on the softening length} 
	\label{sec:softening}
	
\begin{figure} 
  \centering
    \includegraphics[width=\columnwidth]{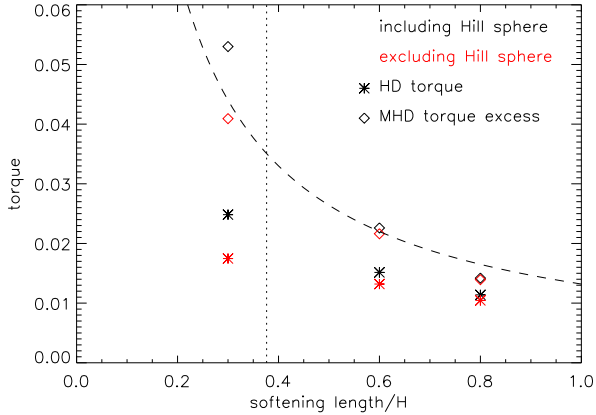}
  \caption{MHD torque excess (diamonds) and hydrodynamical torque (stars) as a function of the softening length of the planet's gravitational potential. The simulations were performed using Model~2 and the fiducial parameters. Red symbols correspond to torques computed excluding the Hill sphere of the planet, while it is not excluded in the torques shown by the black symbols. The dashed and dash-dotted lines show a dependence as $1/\epsilon$ for comparison. The vertical dotted line shows the value of the planet's Hill radius (given by equation~\ref{eq:hill}). }
    	\label{fig:torque_eps}
\end{figure}

So far all the results presented in this paper have used a softening $\epsilon=0.6H(r_p)$ for the gravitational potential of the planet. It is not clear which value of the softening length is the best representation of the 3D behaviour in discs, therefore it is useful to study the dependence of our results with this parameter. Figure~\ref{fig:torque_eps} shows the dependence of the hydrodynamical torque and of the MHD torque excess on the softening length. As the softening length is decreased, both torques increase significantly. The MHD torque excess has a dependence that is roughly consistent with being inversely proportional to $\epsilon$ in this range of softening length, comparable to the scaling of the hydrodynamical corotation torque in locally isothermal discs \citep{pbck10}. We also considered the effect of excluding the planet's Hill sphere from the torque calculation (which was not done so far in this paper). This has a significant impact only if the softening length is smaller than the Hill radius, and therefore barely affects the results for our fiducial softening length.

Another point of interest is the behaviour of the stagnation point as the softening length is decreased. Indeed, for small enough values of the softening length there may actually be three stagnation points in the hydrodynamical case \citep{cm09}. Given the importance of the stagnation point in our interpretation of the MHD torque excess, we may wonder how such a behaviour may affect our results. In the case of a softening length of $\epsilon=0.3H$, we do observe the appearance of three stagnation points in our hydrodynamical simulations in agreement with the results of \citet{cm09}. However, the corresponding MHD simulation still shows only one stagnation point and a very similar topology as the simulations with a larger softening length. The sign of the stagnation point azimuth is then that of the outer X-point (as defined by \citet{cm09} as the intersection of the outer separatrices, which delimit the horseshoe region). The MHD dynamics therefore seems to depend on the azimuthal asymmetry of the hydrodynamical situation but not on the details of the topology of the streamlines.

\section{Comparison with 3D MHD simulations of discs with turbulence due to the MRI}
	\label{sec:turblike}

\begin{figure*} 
  \centering
   \includegraphics[width=\columnwidth]{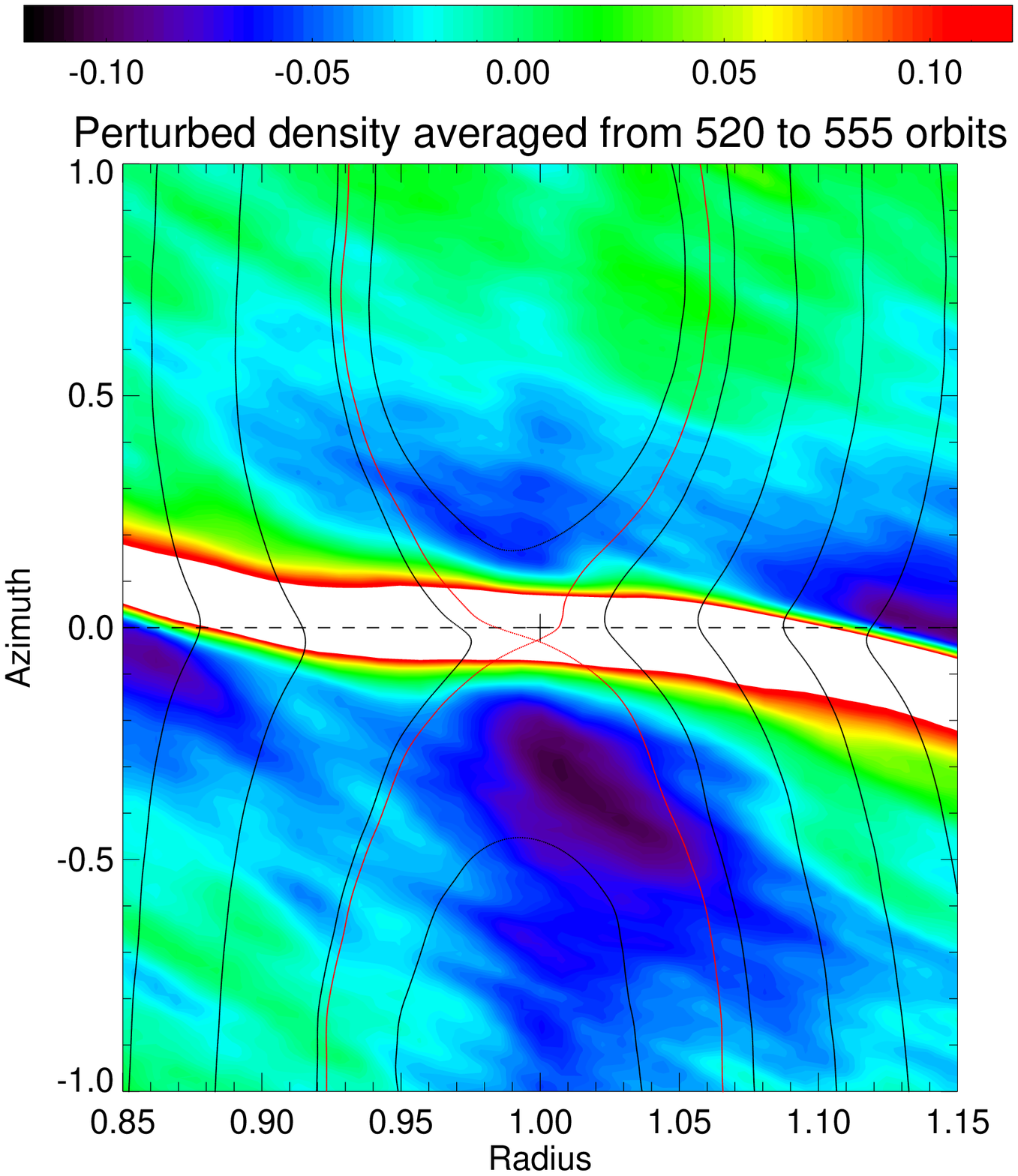}
   \includegraphics[width=0.927\columnwidth]{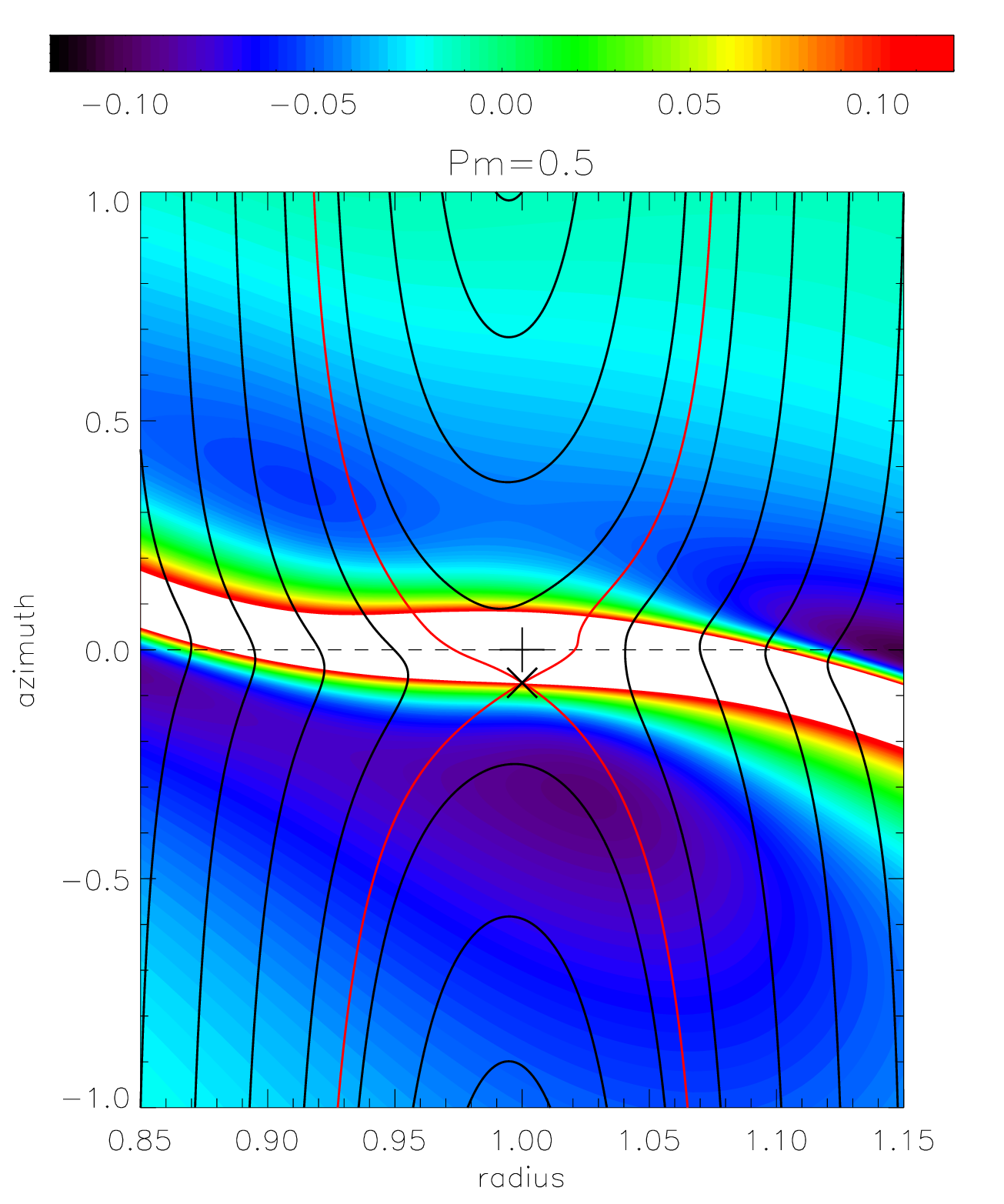}	
   \caption{Surface density perturbation: comparison of the 3D turbulent simulations of  BFNM11 (left panel) and our  2D laminar simulations (right panel). The 3D simulations have been averaged in the vertical direction and on a  time-period of 35 orbits between 520 and 555 orbits. The azimuthally averaged surface density has been subtracted to obtain the density perturbations, shown with the colour contours. The streamlines are shown with black lines, and the horseshoe separatrices are represented by red lines. We show the result of Model~1 ($p=-1/2$ and $q=-1$), and the 2D simulations use a magnetic Prandtl number of $\Pm=0.5$ (the result with $\Pm=1$ is qualitatively very similar with only a slightly larger underdensity near the horseshoe separatrix).  }
   	\label{fig:turblike_drho}
\end{figure*}

\begin{figure*} 
  \centering
   \includegraphics[width=\columnwidth]{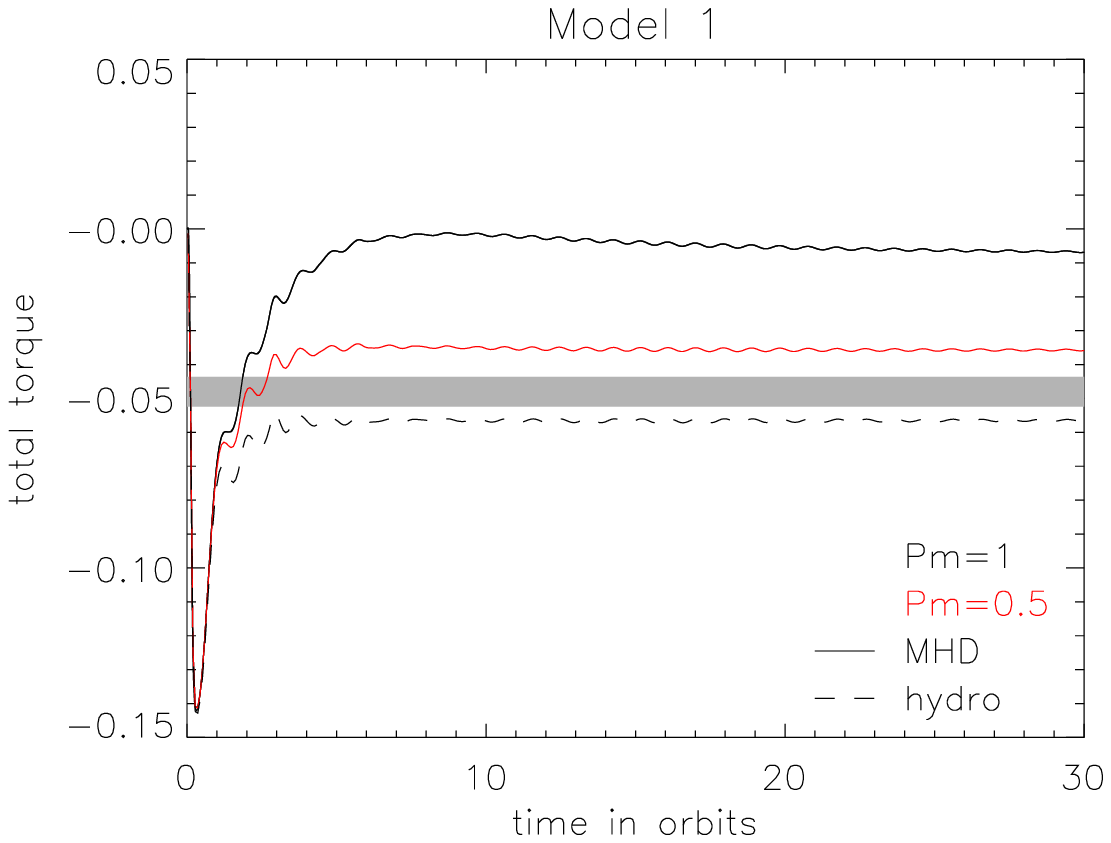}
   \includegraphics[width=\columnwidth]{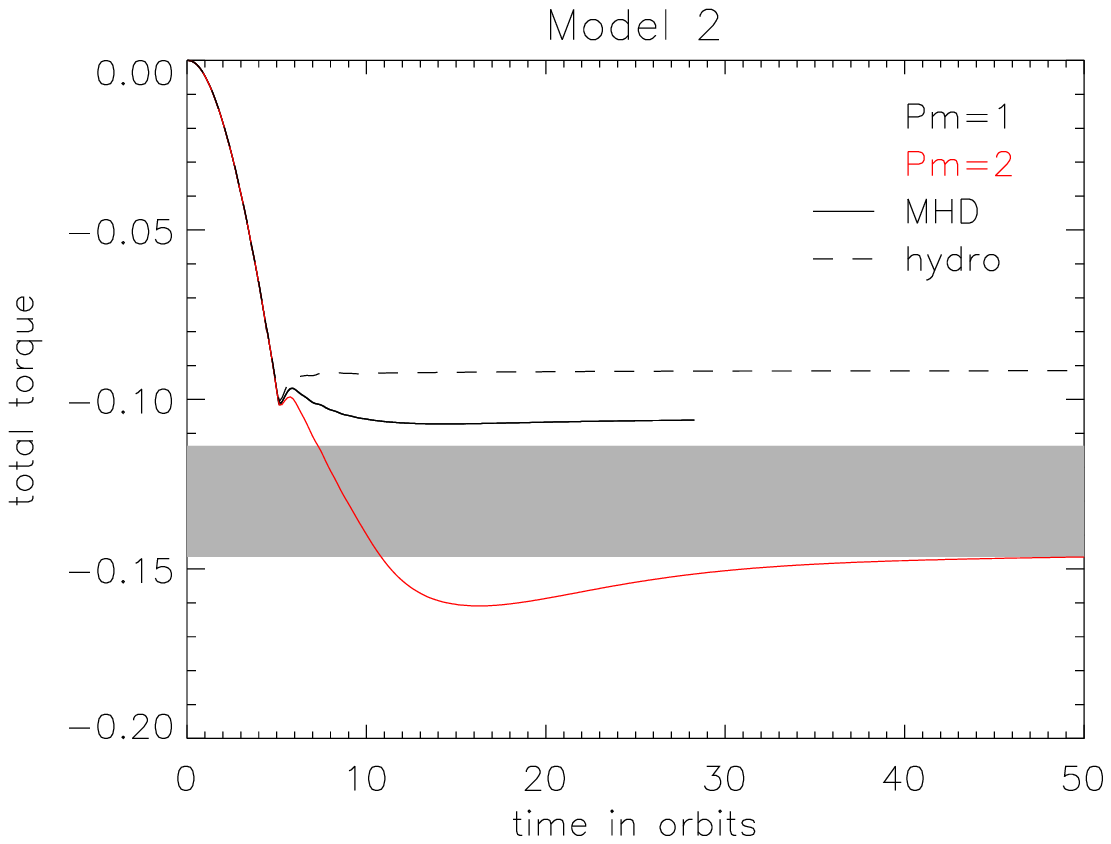}	
  \includegraphics[width=\columnwidth]{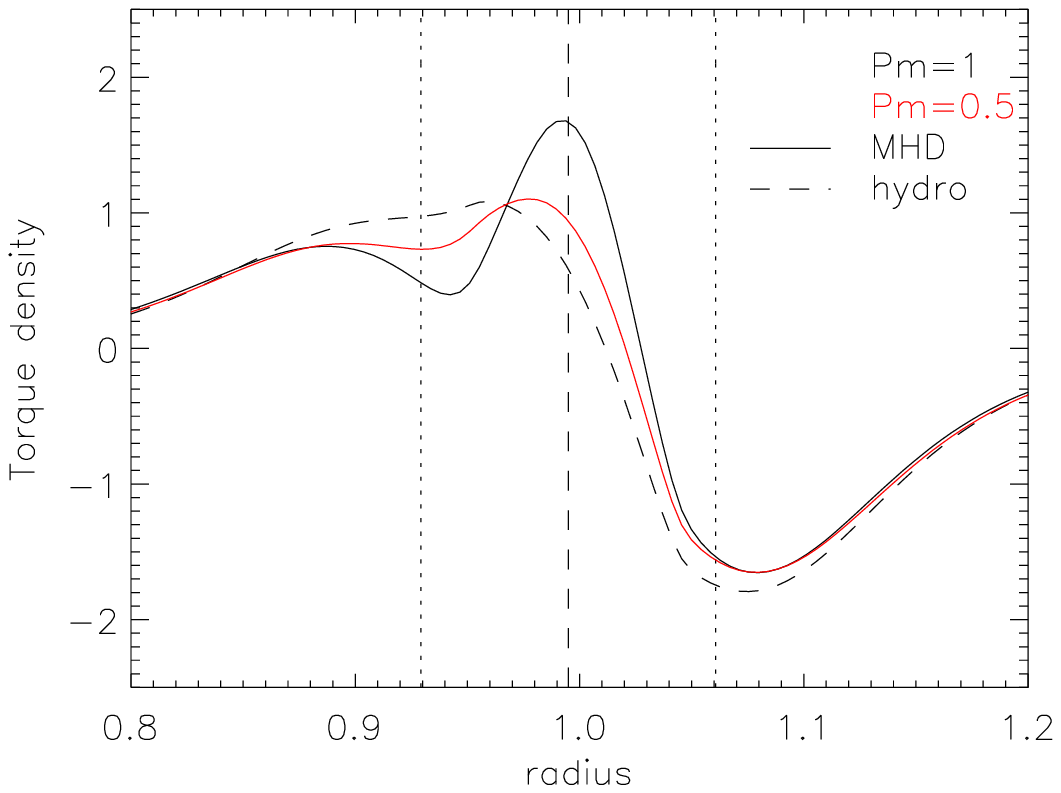}
   \includegraphics[width=\columnwidth]{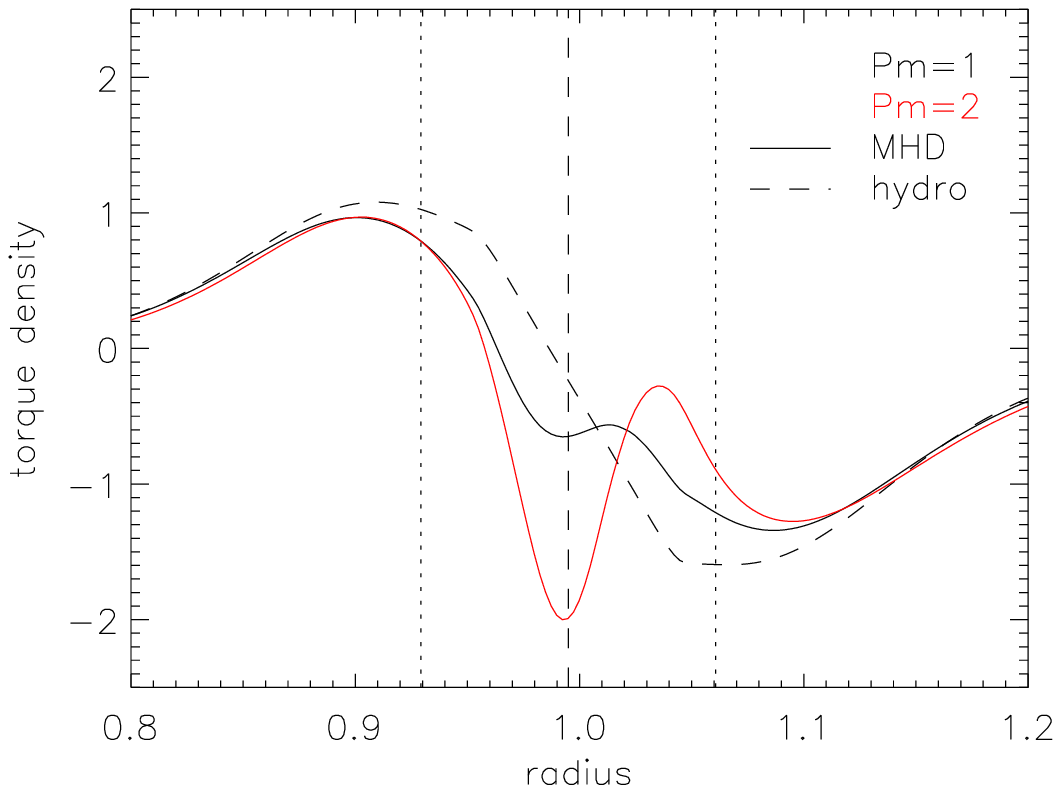}
   \includegraphics[width=\columnwidth]{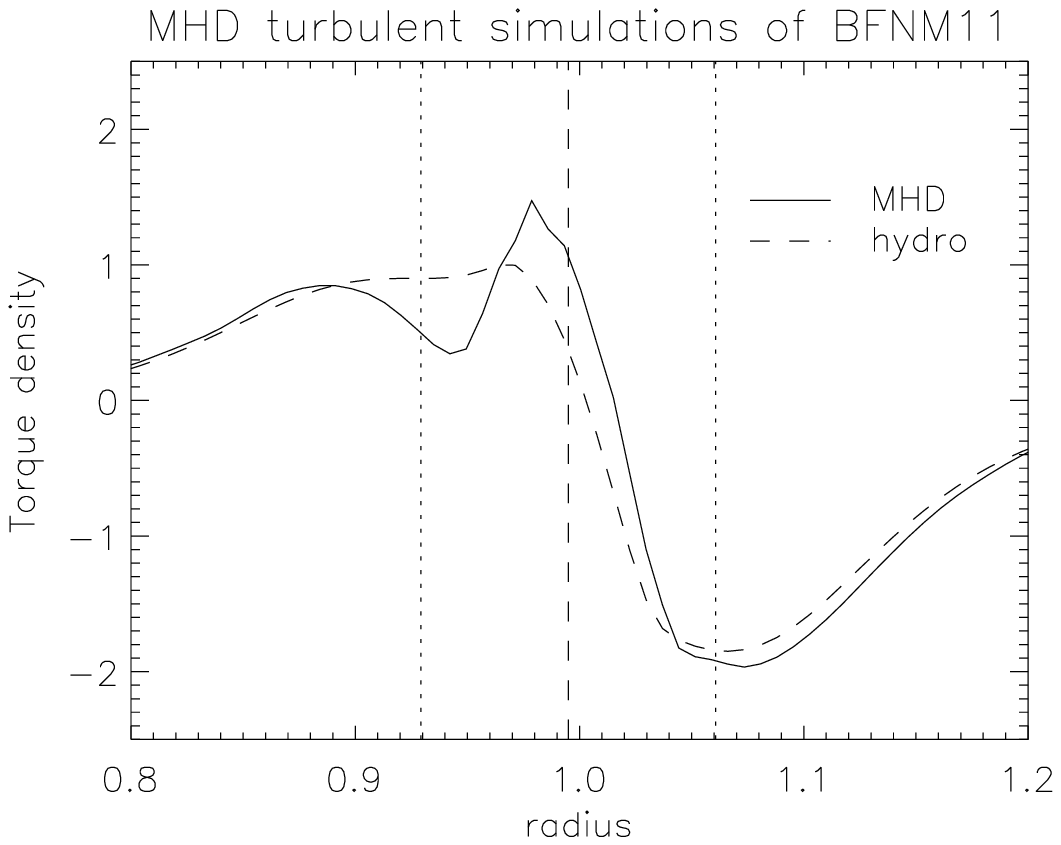}
   \includegraphics[width=\columnwidth]{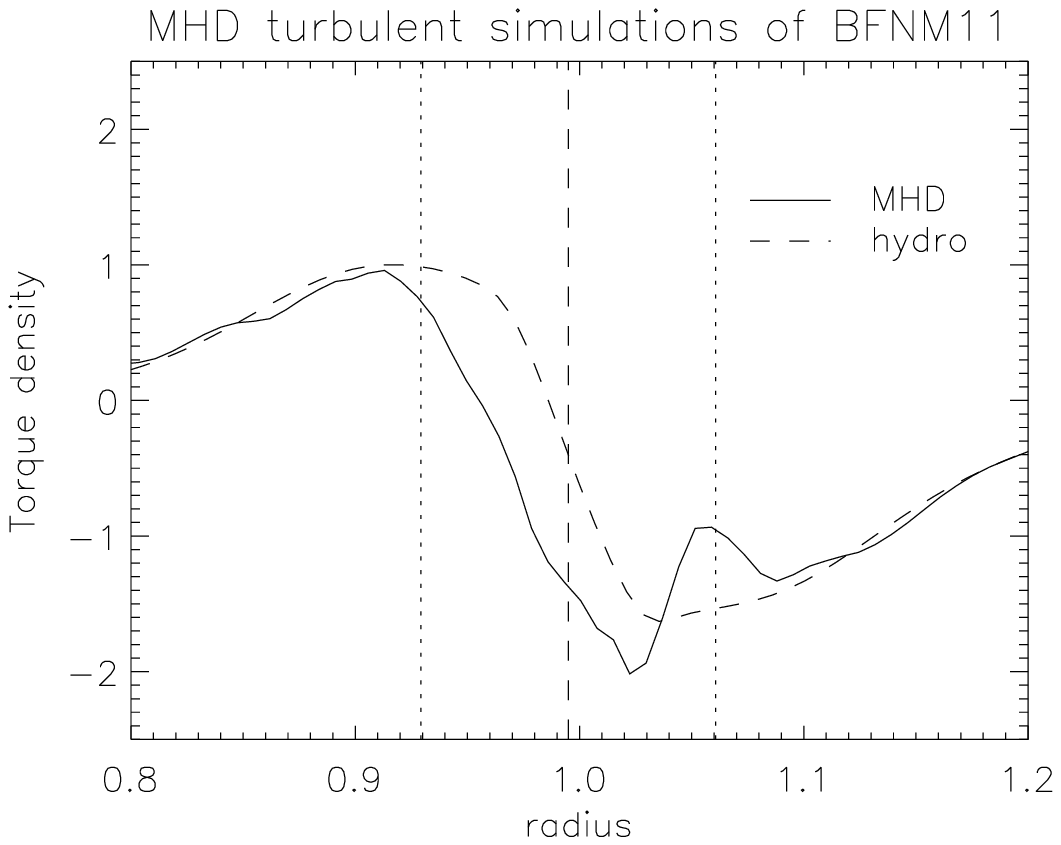}
   \caption{Comparison of our 2D laminar simulations with the 3D turbulent simulations of Baruteau et al. (2011, BFNM11). Two disc models are considered: Model 1, i.e. $p=-1/2$ and $q=-1$ (left column, simulations performed with RAMSES), and Model 2, i.e. $p=-3/2$, $q=0$ (right column, simulations performed with NIRVANA). Upper row: total torque as a function of time. The range of steady state torque values obtained in the MHD turbulent simulations of BFNM11 is illustrated with a grey shaded area. Middle row: torque density distribution in the steady state in our laminar simulations. Lower row : torque density distribution obtained by BFNM11 with the code NIRVANA (the radius has been normalised by the orbital radius of the planet, and the torque density by the maximum of the hydrodynamical density distribution). In all panels hydrodynamic simulations are shown with dashed lines, while their MHD counterpart are shown with full lines. The viscosity and magnetic field strength are set to $\alpha=3.10^{-2}$, and $\beta=50$ as measured in BFNM11. Furthermore, we considered 2 values of the magnetic Prandtl number (setting the resistivity) : $P_m=1$ (black lines) and $P_m=0.5$ (red lines). The torques and torque densities are calculated excluding the Hill radius for consistency with BFNM11. }
   	\label{fig:turblike}
\end{figure*}

In this section we compare our results with the 3D MHD simulations of discs performed by  BFNM11, where the MRI operates throughout the whole disc. The goal is to check how well a 2D description where the effects of turbulence are modelled by diffusion coefficients can explain a 3D turbulent situation. For comparison purposes, we ran 2D simulations using the same parameters as in BFNM11. We study two disc models differing in the slope of the temperature and surface density profiles: model 1 ($p=-1/2$ and $q=-1$) and model 2 ($p=-3/2$ and $q=0$) as defined earlier in this paper. The other parameters are: $M_p=3\times10^{-4}$, $H/r = 0.1$, as well as $\beta=50$ and $\alpha=3\times10^{-2}$ as measured in the turbulent simulations. The simulations of BFNM11 are not vertically stratified and use a softening length of the planet potential $\epsilon=0.2H(r_p)$, which we adopt here. With this rather small value of the softening length, the Hill sphere can have an effect on the torque (see Section~\ref{sec:softening}). For consistency with BFNM11, we exclude the Hill sphere from the torque calculations. In a 3D stratified simulation, the structure of the flow may differ somewhat from that obtained with a softened potential. This should be considered in the future for a better quantitative estimate of the MHD torque excess. The slope of the magnetic field strength, which was found to have no effect on the MHD torque excess (see Section~\ref{sec:torque_b}), was given our fiducial value $b=-1$, which avoids diffusion since it is current-free. The effective magnetic Prandtl number was not measured in BFNM11 because it is not a straightforward analysis and requires to run specific simulations. As a result, the magnetic Prandtl number needs to be specified based on expectations from the literature. As discussed in Section~\ref{sec:setup_disc}, it is expected to be of order unity but may differ from it by a factor of a few. We therefore considered three values of this parameter in order to determine its influence on the results : $\Pm=2$, $\Pm=1$, and $\Pm=0.5$. In order to avoid a viscous evolution of the surface density profile, we used a radial profile of the viscosity such that the radial velocity vanishes (a uniform value of the diffusion coefficients in model 1, and an $\alpha$ parameter scaling like $r^{-1/2}$ in model 2). In addition to the MHD simulations, we also ran hydrodynamical simulations with the same parameters. 

Figure~\ref{fig:turblike_drho} illustrates the overall flow topology of Model~1 in turbulent and laminar simulations, by showing the density perturbation and the streamlines. The width of the horseshoe region is in good agreement between the turbulent and the laminar simulation. In both simulations, we clearly observe an underdense lobe near the downstream separatrix at negative azimuth. A much weaker underdensity is also marginally seen at positive azimuth. We showed in Section~\ref{sec:magnetic} and \ref{sec:torque_model1-2} that these underdense lobes are due to an accumulation of magnetic field near the downstream separatrix of the horseshoe region, and that the azimuthal asymmetry of this underdensity (clearly seen here) leads to a net torque on the planet. Importantly, the stagnation point is located at a negative azimuth in both the turbulent and the laminar simulations, confirming that the most prominent underdense lobe lies on the same side of the planet as the stagnation point. This comparison shows that the overall flow topology and the density perturbation is in good agreement between the turbulent simulations of BFNM11 and our laminar simulations.

The results of the torque calculations are illustrated by Figure~\ref{fig:turblike}, showing the time evolution of the torque (upper row) and the torque density distribution (middle row for our simulations, lower row for the turbulent simulations of BFNM11). The range of values of the total torque in steady state obtained in the MHD turbulent simulations of BFNM11 is shown in the upper panels with a grey shaded area. Several features show a very nice qualitative agreement between our laminar simulations and the turbulent simulations of BFNM11. First the sign of the MHD torque excess: as earlier in this paper we find a positive torque excess in Model~1 and a negative one in Model~2. This agrees with the results of BFNM11 who find a marginally positive torque excess in Model~1 and a negative one in Model~2. Secondly the shape of the torque density distribution shows two additional features in the presence of a magnetic field as compared to the hydrodynamical simulation: a negative contribution in the inner part of the horseshoe region and a positive one in the outer part, as was already observed by BFNM11. These are due respectively to the underdense lobes in front and behind the planet, which are located along the downstream separatrices where the magnetic flux accumulates (see Sections~\ref{sec:magnetic} and \ref{sec:torque_model1-2}). We also observe that the two peaks are not symmetric about the planet radius or the corotation radius. They are shifted to smaller radii in Model~1 and to larger radii in Model~2, again in agreement with BFNM11. This asymmetry can be related to the azimuthal shift of the stagnation point with respect to the planet, since the underdense region lies along the downstream separatrices. In Model~1 the stagnation point azimuth is negative, such that the underdense region lies behind the planet on a larger proportion of the horseshoe region and therefore the positive torque contribution extends to radii smaller than the corotation radius. Conversely, in Model~2 the stagnation point azimuth is positive, the underdense lobe is preferentially in front of the planet, and the negative torque contribution extends to radii larger than the corotation radius. The qualitative agreement (in the torque excess, the torque density distribution, the surface density perturbations, and the azimuth of the stagnation point) gives confidence that the physical picture described in this paper applies to the turbulent simulations of BFNM11.

Quantitatively, however, the agreement between the torque calculations in laminar and turbulent simulations is not so good, though the comparison is made difficult by the strong dependence on the magnetic Prandtl number. Starting the comparison with the fiducial value $\Pm=1$, we observe that the amplitude of the MHD torque excess is too large in Model~1 and too small in Model~2 as compared to BFNM11. This discrepancy can be resolved by choosing the right magnetic Prandtl number, however a different value is needed in both models: $\Pm \lesssim 0.5 $ in Model~1, and $\Pm\simeq1-2$ in Model~2. This is not entirely satisfactory at the present because we can give no reason for the magnetic Prandtl number of the MHD turbulence to be different in the two disc models. However, this may indicate a sensitivity of the results to the way turbulent diffusion operates and this may differ in the two cases. We note that the range of magnetic Prandtl numbers needed to explain the results of BFNM11 is in line with numerical simulations of turbulence driven by the MRI, where the effective magnetic Prandtl number is measured to be in the range $\Pm \sim [0.3-5]$ depending on the magnetic field configuration  \citep{guan09,lesur09,fromang09}.

Comparing the laminar simulations presented in this section to the results of the previous sections, an unexpected feature appears: the amplitude of the MHD torque excess differs by a factor slightly more than $3$ between the two disc models, while it was found in Section~\ref{sec:torque_gradients} that the MHD torque excess barely depends on the density and temperature profiles. This different behaviour is probably caused by non-linear effects, which are much more pronounced here due to the larger planet's mass ($M_p/(M_*h^3)=0.3$) and the smaller softening length. Indeed, we checked that increasing the softening length to $\epsilon=0.6H$ reduced the discrepancy between the two models (a factor $2$ in the MHD torque excess). Also, we checked in Figure~\ref{fig:torque_alpha} that the result of Section~\ref{sec:torque_gradients} was not a peculiarity of the fiducial parameters:  for several values of $\alpha$, models~1 and 2 have similar amplitudes of the MHD torque excess. Therefore, we expect the results presented in Section~\ref{sec:torque} to be representative of the migration of low mass planets when the interaction of the planet with the disc is linear. Additional complexity may arise when non-linear effects become important at larger planet masses and smaller softening lengths. 

\citet{Uribe11} also performed simulations of planet migration in 3D MHD turbulent discs. We do not attempt a detailed comparison with their results because the disc density profile and the strength of the turbulence significantly evolve during the time of their simulations. A qualitative comparison is however interesting. In their simulations of intermediate planet mass experiencing type I migration (with $M_p/(M_*h^3)\simeq 0.15-0.6$), they observed a positive torque exerted on the planet, corresponding to outward migration. They interpreted this result as a consequence of a strong and positive hydrodynamical corotation torque in a locally flat or increasing surface density profile as described by \citet{masset06a}. In light of our results, we suggest that the MHD torque excess might have played a prominent role in the positive torque they observed. Indeed, the density and temperature profiles in their simulations (initially the same as Model~1, and later on showing a locally flat or increasing density profile) should lead to a positive MHD torque excess (see Figure~\ref{fig:phistagn_hydro} or equation~\ref{eq:phistagn_sign}). It is hard to make any prediction for the amplitude of the MHD torque excess given that they used different planet masses and disc aspect ratio than our fiducial values, but we note that the low value of the effective viscosity measured in their simulations ($\alpha\simeq 1-2\times10^{-3}$) is favourable for a strong MHD torque excess.

\section{Discussion and conclusion}
	\label{sec:conclusion}
\subsection{Summary}
	\label{sec:summary}
	
Using two-dimensional numerical simulations of viscous resistive laminar discs, we have confirmed and explained the existence of an additional corotation torque due a weak azimuthal magnetic field, which was observed in the 3D MHD turbulent disc simulations by BFNM11. It arises from the interaction between the horseshoe motion of the gas in the planet's coorbital region and an azimuthal magnetic field in the protoplanetary disc. We refer to this new torque component as the MHD torque excess. This is distinct from the magnetic resonances described by \citet{terquem03}, and arises in a different regime of magnetic field strength. Indeed, we considered magnetic fields that are weak enough not to prevent the horseshoe dynamics. This is the case in our simulations where the Alfv\'en speed is smaller than the shear velocity at the separatrix of the horseshoe region. We do not see any evidence of the presence of magnetic resonances in this regime. As discussed in Section~\ref{sec:fiducial}, this may be due either to the horseshoe dynamics (the presence of the horseshoe trajectories radically changes the shear, such that the description in terms of magnetic resonances is not appropriate anymore) or the effect of resistive diffusion, which we included to describe the effect of turbulence.

The physical origin of this new component of the corotation torque has been interpreted in the following way. The horseshoe motion of the gas near the planet tends to accumulate the magnetic flux contained in the horseshoe region on the downstream separatrix. This results in a  region of enhanced magnetic pressure around the downstream separatrices, where the gas surface density is reduced to maintain approximate total pressure balance. The MHD torque excess most likely stem from an azimuthal asymmetry of this underdense region. This is linked with an azimuthal shift of the stagnation point with respect to the planet, since the underdensity is maximum close to the stagnation point. Such an asymmetry is always observed in the presence of a magnetic field, and increases with the magnetic field strength. A positive azimuth of the stagnation point leads to a negative torque excess, while a negative azimuth of the stagnation point leads to a positive torque excess. This sign is governed by a hydrodynamic process determined by the temperature and density gradients. The amplitude of the torque excess on the other hand is governed by a MHD process determined by the magnetic field strength and diffusivity.

The properties of the MHD torque excess may be summarised in the following way:
\begin{itemize}
\item Its magnitude is roughly independent of the gradients of magnetic field strength, density and temperature, which have at most an effect of order $10-20\%$. In the non-dimensional problem, it therefore depends only on the magnetic field strength (through $\beta$), the diffusion coefficients ($\alpha$ and $\Pm$), the planet-to-primary mass ratio $M_p/M_*$, the disc's aspect ratio $h$, and the ratio of softening length of the planet gravitational potential to the pressure scaleheight $\epsilon/H$. By varying the parameters $\alpha$ and $\beta$, we obtained the following scaling of the amplitude of the MHD torque excess: $\Gamma\propto 1/(\beta^2\alpha^4)$. Thus it increases very strongly when the magnetic strength is increased or when the turbulent resistivity is decreased. The amplitude is also roughly consistent with being inversely proportional to the softening length of the planet potential. The dependence on $M_p/M_*$, and $h$ is left for future work. 

\item The sign of the MHD torque excess is determined by whether the azimuth of the stagnation point is located ahead of or behind the planet, and this depends solely on the density and temperature gradients at the planet's location. It is therefore independent of the strength and gradient of magnetic field and of the viscosity and resistivity. Our numerical results show that the sign of the MHD torque is determined by the slope indices of the temperature profile $q$, and surface density profile $p$ in the following manner:
\begin{equation}
{\rm sign}(\Gamma) = {\rm sign}(p-1.3q+0.6)
\end{equation}
\end{itemize}

We compared the results of our 2D laminar simulations with the 3D turbulent simulations of BFNM11. We found good qualitative agreement for the sign of the MHD torque excess and for the torque density distribution, and their dependence on the density and temperature profiles. The quantitative comparison of the MHD torque excess amplitude is complicated by its steep dependence on the effective resistivity. The turbulent simulations results can be reproduced with values of the magnetic Prandtl number ranging from 0.5 to 2 depending on the disc model (density and temperature profiles).

\subsection{Discussion}
	\label{sec:discussion}
We have estimated the MHD torque excess by performing  non linear numerical simulations and found that it can have a significant impact on the migration. 
The  amplitude of the MHD torque excess  was found to be a very steep function of the magnetic field strength and diffusivity (through $\alpha$ and $\beta$), which are rather uncertain. The magnetic field strength in protoplanetary discs is so far poorly constrained observationally, but the SPIRou spectropolarimeter might provide more information in the future. From a theoretical perspective, numerical simulations of MHD turbulent discs suggest values of $\alpha$ between a few times $10^{-3}$ and a few times $10^{-2}$ in the absence of a net vertical magnetic flux \citep{fromang07b}. This range of values is also compatible with the observed characteristic lifetime of protoplanetary discs of a few millions years. The fiducial parameters used in this paper ($\alpha=5.10^{-3}$, $\beta=100$) are thus plausible for MRI active regions of protoplanetary discs, and they lead to a strong MHD torque excess that is able to reverse the direction of migration. This suggests that the migration of low mass planets in MRI active regions of protoplanetary discs could be significantly affected by the magnetic field. This is further suggested by the comparison with MHD turbulent simulations of planet migration detailed in Section~\ref{sec:turblike}, although further studies are certainly necessary to better characterise the influence of the MHD torque excess in self-consistent simulations with MHD turbulence.

Could the MHD torque excess also be significant in dead zones, where the ionisation is too low for MHD turbulence to be sustained ? This possibility may seem counterintuitive at first sight because dead zones are characterised by a large physical resistivity that prevents turbulence, while a significant MHD torque excess needs the resistivity to be low enough. However both criteria may be satisfied at the same time. Indeed the critical magnetic Reynolds number below which the turbulence is shut off is thought to lie in the range $10^2-10^4$ depending on the presence or not of a net vertical magnetic flux \citep{fleming00}, which corresponds to $\alpha/\Pm = 10^{-4}-10^{-2}$. Our results show that the MHD torque excess can be significant for this range of resistivity if an azimuthal magnetic field is present in the dead zone. The simulations of \citet{turner08} suggest that such an azimuthal magnetic field can arise in a dead zone owing to the interaction with the active surface layers of the disc. The presence of a significant MHD torque excess taking place in dead zones is therefore an intriguing possibility which should be further studied. 

What sign of the MHD torque excess should one expect for disc profiles typical of protoplanetary discs ? Answering this question is difficult because the structure of protoplanetary discs is still poorly known especially in their inner parts. Due to angular resolution limitations, observations can so far only constrain the outer parts of protoplanetary discs at distances of several tens of AU. In this disc region, they suggest the following temperature and density profiles: $q \in [-0.7,-0.4]$, $p\in[-1,0]$ \citep[and references therein]{williams11}. This region of parameter space lies entirely in the region where the MHD torque excess is positive (see Figure~\ref{fig:phistagn_hydro}). We therefore expect that in outer regions of protoplanetary discs, the MHD torque excess can slow down or reverse an otherwise inwards type I migration. The density and temperature profiles in the inner part of the disc are much more uncertain so far, though the ALMA interferometer should enable its study in the near future. One might speculate that very negative values of $p$ are unlikely and therefore that the MHD torque excess is likely positive. We should caution that all this discussion is assuming smooth density and temperature profiles. However what really matters is the local gradient near the planet in a region of radial width $\sim H$, which may be very different from the global ones in some special locations. For example, a locally flat or positive surface density profile has been argued to provide a planet trap thanks to a strong positive corotation torque \citep{masset06a}. We note that the MHD torque excess is positive for such a density profile (unless the temperature is strongly increasing at the same location) and could thus make the planet trap mechanism even more effective.

As discussed above, an interesting consequence of the MHD torque excess is that it could slow down or reverse an otherwise inward type I migration. This is especially interesting in the outer parts of the disc because they are radiatively efficient (typically at $r \gtrsim 15 \, {\rm AU}$ \citep{pm08}), so that the hydrodynamical corotation torque is probably small and thus inefficient at counteracting the inward migration. Reversing the migration at large radii could be important to explain the observations  by direct imaging of planets at several tens of AU from their star like the 4 planets system HR 8799 \citep{marois08,marois10}. We speculate that the MHD torque excess could lead to an outward migration to large distances and/or might in some cases prevent the inward migration of planets formed at large distances, which were shown to migrate inwards very fast by \citet{bmp11}. As shown in this article, the positive torque leading to outward migration can be as large as the differential Lindblad torque. Using the standard expression of \citet{Tanaka2002} for the differential Lindblad torque in 3D, and a minimum mass solar nebula disc model, we find that a 10 Earth mass planet could migrate to $50\,{\rm AU}$ in $1.5\times 10^6 \, {\rm years}$. There is therefore enough time for the migration to occur during the lifetime of the protoplanetary disc. That the MHD torque excess could lead to a sustained outward migration over long distances remains, however, to be proven. Indeed, as the planet migrates outwards the ratio of horseshoe U-turn time to the diffusion time across the horseshoe region is likely to vary, which would change the efficiency of the MHD torque excess. In this respect, it would be important to investigate the dependence of the MHD torque excess on the disc aspect ratio and the planet mass.

Finally, we stress that we have considered in this study only a simple magnetic configuration: a large scale purely azimuthal magnetic field. In reality, the field geometry is likely to be much more complex. For turbulent regions of the disc, it should be checked further if the effects of a disordered field structure can be described by a mean field with diffusion coefficients, as is assumed here. Another question is the effect of a vertical magnetic field. The strength of the vertical magnetic field in accretion discs is still debated. It depends on the relative efficiency of the advection by the accretion flow and the diffusion by the turbulent resistivity, which is still an active subject of research \citep[and references therein]{guilet12}. The presence of jets in many T-Tauri stars may suggest that a strong vertical magnetic field threads at least the inner part of protoplanetary discs according to the magneto-centrifugal model of jet launched from an accretion disc \citep{blandford82,ferreira06}. It would thus also be important to investigate the influence of a vertical magnetic field on planetary migration, as started by \citet{muto08}.

\section*{Acknowledgments}
This research was supported by STFC. CB is supported by a Herschel Smith Postdoctoral Fellowship of the University of Cambridge. We are grateful to S\'ebastien Fromang, Richard Nelson and Sijme-Jan Paardekooper for useful discussions.

\section*{Appendix : convergence test and code comparison}

We have tested the convergence and compared the results with the two codes used in this article, RAMSES and NIRVANA. For this purpose we chose to use Model~2 with the fiducial parameters both in MHD ($\beta=100$) and in hydrodynamics cases. We used three different resolutions: the nominal resolution used throughout this paper where the half width of the horseshoe region is resolved by about 9 cells (NR, blue curves), a twice lower resolution (LR, black curves) and a twice higher resolution (HR, red curves). Noticeable differences are observed at low resolution although the numerical error on the torque remains modest ($5-10\%$): in particular the two peaks in the torque density distribution are slightly less pronounced than at higher resolution. The agreement between the high and nominal resolution is remarkable with the code RAMSES, as the torques and torque density distributions are almost indistinguishable. This agreement is slightly less good for the code NIRVANA but still quite acceptable with a $5\%$ difference in the torque. Finally the results of the two codes agree reasonable well with each other, with a difference of $\simeq 5\%$ in the torques computed at nominal or high resolutions.

\bibliography{migration}

\begin{thebibliography}{}

\bibitem[\protect\citeauthoryear{{Baruteau}, {Fromang}, {Nelson} \&
  {Masset}}{{Baruteau} et~al.}{2011}]{bfnm11}
{Baruteau} C.,  {Fromang} S.,  {Nelson} R.~P.,    {Masset} F.,  2011, \aap,
  533, A84

\bibitem[\protect\citeauthoryear{{Baruteau} \& {Lin}}{{Baruteau} \&
  {Lin}}{2010}]{bl10}
{Baruteau} C.,  {Lin} D.~N.~C.,  2010, \apj, 709, 759

\bibitem[\protect\citeauthoryear{{Baruteau} \& {Masset}}{{Baruteau} \&
  {Masset}}{2008}]{bm08a}
{Baruteau} C.,  {Masset} F.,  2008, \apj, 672, 1054

\bibitem[\protect\citeauthoryear{{Baruteau}, {Meru} \&
  {Paardekooper}}{{Baruteau} et~al.}{2011}]{bmp11}
{Baruteau} C.,  {Meru} F.,    {Paardekooper} S.-J.,  2011, \mnras, 416, 1971

\bibitem[\protect\citeauthoryear{{Blandford} \& {Payne}}{{Blandford} \&
  {Payne}}{1982}]{blandford82}
{Blandford} R.~D.,  {Payne} D.~G.,  1982, \mnras, 199, 883

\bibitem[\protect\citeauthoryear{{Casoli} \& {Masset}}{{Casoli} \&
  {Masset}}{2009}]{cm09}
{Casoli} J.,  {Masset} F.~S.,  2009, \apj, 703, 845

\bibitem[\protect\citeauthoryear{{Ferreira}, {Dougados} \& {Cabrit}}{{Ferreira}
  et~al.}{2006}]{ferreira06}
{Ferreira} J.,  {Dougados} C.,    {Cabrit} S.,  2006, \aap, 453, 785

\bibitem[\protect\citeauthoryear{{Ferreira} \& {Pelletier}}{{Ferreira} \&
  {Pelletier}}{1995}]{ferreira95}
{Ferreira} J.,  {Pelletier} G.,  1995, \aap, 295, 807

\bibitem[\protect\citeauthoryear{{Fleming}, {Stone} \& {Hawley}}{{Fleming}
  et~al.}{2000}]{fleming00}
{Fleming} T.~P.,  {Stone} J.~M.,    {Hawley} J.~F.,  2000, \apj, 530, 464

\bibitem[\protect\citeauthoryear{{Fromang}, {Hennebelle} \&
  {Teyssier}}{{Fromang} et~al.}{2006}]{fromang06}
{Fromang} S.,  {Hennebelle} P.,    {Teyssier} R.,  2006, \aap, 457, 371

\bibitem[\protect\citeauthoryear{{Fromang}, {Papaloizou}, {Lesur} \&
  {Heinemann}}{{Fromang} et~al.}{2007}]{fromang07b}
{Fromang} S.,  {Papaloizou} J.,  {Lesur} G.,    {Heinemann} T.,  2007, \aap,
  476, 1123

\bibitem[\protect\citeauthoryear{{Fromang} \& {Stone}}{{Fromang} \&
  {Stone}}{2009}]{fromang09}
{Fromang} S.,  {Stone} J.~M.,  2009, \aap, 507, 19

\bibitem[\protect\citeauthoryear{{Fromang}, {Terquem} \& {Nelson}}{{Fromang}
  et~al.}{2005}]{fromang05}
{Fromang} S.,  {Terquem} C.,    {Nelson} R.~P.,  2005, \mnras, 363, 943

\bibitem[\protect\citeauthoryear{{Guan} \& {Gammie}}{{Guan} \&
  {Gammie}}{2009}]{guan09}
{Guan} X.,  {Gammie} C.~F.,  2009, \apj, 697, 1901

\bibitem[\protect\citeauthoryear{{Guilet} \& {Ogilvie}}{{Guilet} \&
  {Ogilvie}}{2012}]{guilet12}
{Guilet} J.,  {Ogilvie} G.~I.,  2012, \mnras, 424, 2097

\bibitem[\protect\citeauthoryear{{Hellary} \& {Nelson}}{{Hellary} \&
  {Nelson}}{2012}]{HellaryNelson12}
{Hellary} P.,  {Nelson} R.~P.,  2012, \mnras, 419, 2737

\bibitem[\protect\citeauthoryear{{Ida} \& {Lin}}{{Ida} \&
  {Lin}}{2008}]{IdaLin4}
{Ida} S.,  {Lin} D.~N.~C.,  2008, \apj, 673, 487

\bibitem[\protect\citeauthoryear{{Kley}, {Bitsch} \& {Klahr}}{{Kley}
  et~al.}{2009}]{Kley09}
{Kley} W.,  {Bitsch} B.,    {Klahr} H.,  2009, \aap, 506, 971

\bibitem[\protect\citeauthoryear{{Laughlin}, {Steinacker} \&
  {Adams}}{{Laughlin} et~al.}{2004}]{lsa04}
{Laughlin} G.,  {Steinacker} A.,    {Adams} F.~C.,  2004, \apj, 608, 489

\bibitem[\protect\citeauthoryear{{Lesur} \& {Longaretti}}{{Lesur} \&
  {Longaretti}}{2009}]{lesur09}
{Lesur} G.,  {Longaretti} P.-Y.,  2009, \aap, 504, 309

\bibitem[\protect\citeauthoryear{{Lin} \& {Papaloizou}}{{Lin} \&
  {Papaloizou}}{1986}]{lp86}
{Lin} D.~N.~C.,  {Papaloizou} J.,  1986, \apj, 309, 846

\bibitem[\protect\citeauthoryear{{Lubow}, {Papaloizou} \& {Pringle}}{{Lubow}
  et~al.}{1994}]{lubow94a}
{Lubow} S.~H.,  {Papaloizou} J.~C.~B.,    {Pringle} J.~E.,  1994, \mnras, 267,
  235

\bibitem[\protect\citeauthoryear{{Marois}, {Macintosh}, {Barman}, {Zuckerman},
  {Song}, {Patience}, {Lafreni{\`e}re} \& {Doyon}}{{Marois}
  et~al.}{2008}]{marois08}
{Marois} C.,  {Macintosh} B.,  {Barman} T.,  {Zuckerman} B.,  {Song} I.,
  {Patience} J.,  {Lafreni{\`e}re} D.,    {Doyon} R.,  2008, Science, 322, 1348

\bibitem[\protect\citeauthoryear{{Marois}, {Zuckerman}, {Konopacky},
  {Macintosh} \& {Barman}}{{Marois} et~al.}{2010}]{marois10}
{Marois} C.,  {Zuckerman} B.,  {Konopacky} Q.~M.,  {Macintosh} B.,    {Barman}
  T.,  2010, \nat, 468, 1080

\bibitem[\protect\citeauthoryear{{Masset}}{{Masset}}{2000}]{fargo1}
{Masset} F.,  2000, \aaps, 141, 165

\bibitem[\protect\citeauthoryear{{Masset}}{{Masset}}{2001}]{masset01}
{Masset} F.~S.,  2001, \apj, 558, 453

\bibitem[\protect\citeauthoryear{{Masset} \& {Casoli}}{{Masset} \&
  {Casoli}}{2009}]{mc09}
{Masset} F.~S.,  {Casoli} J.,  2009, \apj, 703, 857

\bibitem[\protect\citeauthoryear{{Masset} \& {Casoli}}{{Masset} \&
  {Casoli}}{2010}]{mc10}
{Masset} F.~S.,  {Casoli} J.,  2010, \apj, 723, 1393

\bibitem[\protect\citeauthoryear{{Masset}, {Morbidelli}, {Crida} \&
  {Ferreira}}{{Masset} et~al.}{2006}]{masset06a}
{Masset} F.~S.,  {Morbidelli} A.,  {Crida} A.,    {Ferreira} J.,  2006, \apj,
  642, 478

\bibitem[\protect\citeauthoryear{{Masset} \& {Papaloizou}}{{Masset} \&
  {Papaloizou}}{2003}]{mp03}
{Masset} F.~S.,  {Papaloizou} J.~C.~B.,  2003, \apj, 588, 494

\bibitem[\protect\citeauthoryear{{Mordasini}, {Alibert}, {Benz} \&
  {Naef}}{{Mordasini} et~al.}{2009}]{mordasini09b}
{Mordasini} C.,  {Alibert} Y.,  {Benz} W.,    {Naef} D.,  2009, \aap, 501, 1161

\bibitem[\protect\citeauthoryear{{Muto}, {Machida} \& {Inutsuka}}{{Muto}
  et~al.}{2008}]{muto08}
{Muto} T.,  {Machida} M.~N.,    {Inutsuka} S.-i.,  2008, \apj, 679, 813

\bibitem[\protect\citeauthoryear{{Paardekooper}, {Baruteau}, {Crida} \&
  {Kley}}{{Paardekooper} et~al.}{2010}]{pbck10}
{Paardekooper} S.,  {Baruteau} C.,  {Crida} A.,    {Kley} W.,  2010, \mnras,
  401, 1950

\bibitem[\protect\citeauthoryear{{Paardekooper}, {Baruteau} \&
  {Kley}}{{Paardekooper} et~al.}{2011}]{pbk11}
{Paardekooper} S.,  {Baruteau} C.,    {Kley} W.,  2011, \mnras, 410, 293

\bibitem[\protect\citeauthoryear{{Paardekooper} \& {Mellema}}{{Paardekooper} \&
  {Mellema}}{2006}]{pm06}
{Paardekooper} S.-J.,  {Mellema} G.,  2006, \aap, 459, L17

\bibitem[\protect\citeauthoryear{{Paardekooper} \& {Mellema}}{{Paardekooper} \&
  {Mellema}}{2008}]{pm08}
{Paardekooper} S.-J.,  {Mellema} G.,  2008, \aap, 478, 245

\bibitem[\protect\citeauthoryear{{Paardekooper} \& {Papaloizou}}{{Paardekooper}
  \& {Papaloizou}}{2008}]{pp08}
{Paardekooper} S.-J.,  {Papaloizou} J.~C.~B.,  2008, \aap, 485, 877

\bibitem[\protect\citeauthoryear{{Paardekooper} \& {Papaloizou}}{{Paardekooper}
  \& {Papaloizou}}{2009}]{pp09b}
{Paardekooper} S.-J.,  {Papaloizou} J.~C.~B.,  2009, \mnras, 394, 2297

\bibitem[\protect\citeauthoryear{{Pierens}, {Baruteau} \& {Hersant}}{{Pierens}
  et~al.}{2012}]{pierens12}
{Pierens} A.,  {Baruteau} C.,    {Hersant} F.,  2012, \mnras, 427, 1562

\bibitem[\protect\citeauthoryear{{Schlaufman}, {Lin} \& {Ida}}{{Schlaufman}
  et~al.}{2009}]{sli09}
{Schlaufman} K.~C.,  {Lin} D.~N.~C.,    {Ida} S.,  2009, \apj, 691, 1322

\bibitem[\protect\citeauthoryear{{Shakura} \& {Sunyaev}}{{Shakura} \&
  {Sunyaev}}{1973}]{shakura73}
{Shakura} N.~I.,  {Sunyaev} R.~A.,  1973, \aap, 24, 337

\bibitem[\protect\citeauthoryear{{Stone} \& {Norman}}{{Stone} \&
  {Norman}}{1992}]{stone92}
{Stone} J.~M.,  {Norman} M.~L.,  1992, \apjs, 80, 791

\bibitem[\protect\citeauthoryear{{Tanaka}, {Takeuchi} \& {Ward}}{{Tanaka}
  et~al.}{2002}]{Tanaka2002}
{Tanaka} H.,  {Takeuchi} T.,    {Ward} W.~R.,  2002, \apj, 565, 1257

\bibitem[\protect\citeauthoryear{{Terquem}}{{Terquem}}{2003}]{terquem03}
{Terquem} C.~E.~J.~M.~L.~J.,  2003, \mnras, 341, 1157

\bibitem[\protect\citeauthoryear{{Teyssier}}{{Teyssier}}{2002}]{teyssier02}
{Teyssier} R.,  2002, \aap, 385, 337

\bibitem[\protect\citeauthoryear{{Turner} \& {Sano}}{{Turner} \&
  {Sano}}{2008}]{turner08}
{Turner} N.~J.,  {Sano} T.,  2008, \apjl, 679, L131

\bibitem[\protect\citeauthoryear{{Uribe}, {Klahr}, {Flock} \&
  {Henning}}{{Uribe} et~al.}{2011}]{Uribe11}
{Uribe} A.~L.,  {Klahr} H.,  {Flock} M.,    {Henning} T.,  2011, \apj, 736, 85

\bibitem[\protect\citeauthoryear{{Ward}}{{Ward}}{1991}]{wlpi91}
{Ward} W.~R.,  1991, in Lunar and Planetary Institute Conference Abstracts
  {Horsehoe Orbit Drag}.
pp 1463--+

\bibitem[\protect\citeauthoryear{{Ward}}{{Ward}}{1997}]{w97}
{Ward} W.~R.,  1997, Icarus, 126, 261

\bibitem[\protect\citeauthoryear{{Williams} \& {Cieza}}{{Williams} \&
  {Cieza}}{2011}]{williams11}
{Williams} J.~P.,  {Cieza} L.~A.,  2011, \araa, 49, 67

\bibitem[\protect\citeauthoryear{{Zanni}, {Ferrari}, {Rosner}, {Bodo} \&
  {Massaglia}}{{Zanni} et~al.}{2007}]{zanni07}
{Zanni} C.,  {Ferrari} A.,  {Rosner} R.,  {Bodo} G.,    {Massaglia} S.,  2007,
  \aap, 469, 811

\bibitem[\protect\citeauthoryear{{Ziegler} \& {Yorke}}{{Ziegler} \&
  {Yorke}}{1997}]{ziegler97}
{Ziegler} U.,  {Yorke} H.~W.,  1997, Computer Physics Communications, 101, 54

\end{thebibliography}

\bsp
\label{lastpage}

\end{document}